\documentclass[aps,prb,epsf,twocolumn,showpacs]{revtex4-2}
\usepackage{CJK,amsmath,amssymb,graphics,epsfig,epstopdf,color,verbatim,ulem,braket,tabularx,float}
\usepackage[colorlinks,linkcolor=blue,citecolor=blue,urlcolor=blue]{hyperref}
\usepackage[dvipsnames]{xcolor}
\usepackage[marginal]{footmisc}
\usepackage{graphicx}
\usepackage{booktabs}
\usepackage{subfigure}
\UseRawInputEncoding
\begin{document}
\begin{CJK*}{GBK}{song}
\title{\mbox{Magnetic excitations of a trilayer antiferromagnetic Heisenberg model
} 
}

\author{Lan-Ye He, Xin-Man Ye, and Dao-Xin Yao }

\email{yaodaox@mail.sysu.edu.cn}
\affiliation{
\mbox{Guangdong Provincial Key Laboratory of Magnetoelectric Physics and Devices,}
\mbox{State Key Laboratory of Optoelectronic Materials and Technologies,} 
\mbox{Center for Neutron Science and Technology, School of Physics,}\\
\mbox{Sun Yat-sen University, Guangzhou, 510275, China}
}

\begin{abstract}
We investigate the squared sublattice magnetizations and magnetic excitations of a $S=1/2$ trilayer antiferromagnetic Heisenberg model with interlayer interaction $J_{\bot}$ and intralayer interaction $J_{//}$ by employing stochastic series expansion quantum Monte Carlo (SSE-QMC) and stochastic analytic continuation (SAC) methods. Compared with the bilayer model, the trilayer model has one inner layer and two outer layers. The change in its symmetry can lead to special magnetic excitations. Our study reveals that the maximum of the magnetization of the outer sublattice corresponds to smaller ratio parameter $g={J_{//}}/{J_{\bot}}$, a finding that is verified using the finite-size extrapolation. As $g$ decreases, the excitation spectra gradually evolve from a degenerate magnon mode with continua to low-energy and high-energy branches. Particularly when $g$ is small enough, like 0.02, the high-energy spectrum further splits into characteristic doublon ($\approx J_{\bot}$) and quarton ($\approx 1.5 J_{\bot}$) spectral bands. Moreover, the accuracy of the magnetic excitations is confirmed through the SpinW software package and the dispersion relations derived through the linear spin wave theory. Our results provide an important reference for experiments, which can be directly compared with experimental data from inelastic neutron scattering results to verify and guide the accuracy of experimental detection.
\end{abstract}


\date{\today}
\maketitle
\end{CJK*}

\section{Introduction}

Low-dimensional antiferromagnets with spin $S=1/2$ exhibit a variety of complex phenomena in the field of condensed matter physics as a result of their intense quantum fluctuations~\cite{1mermin1966absence,1deiseroth2006fe3gete2,1lu2022observation}. The complicated interplay between quantum mechanics and magnetic interactions is illustrated by the various kinds of ordered and disordered ground states that these systems can generate~\cite{2kim2019suppression,2lee2021magnetic}. The dimensionality crossover effect~\cite{3wang2019determining,3discala2024elucidating} is one of the most intriguing features of these systems. The transition from one-dimensional (1D) chains~\cite{des1962spin,nagler1991spin} to ladder structures~\cite{reigrotzki1994strong} and two-dimensional (2D) systems~\cite{chen1992elementary} is characterized by variations in dimensionality that result in changes in magnetism. The impact of interlayer coupling must be taken into account as we transition from two-dimensional (2D) systems~\cite{richter2010spin,wang2010low} to three-dimensional (3D) systems, whether it is a finite layering of 2D or a genuine three-dimensional structure. This complicates the issue and has, as a result, been less extensively studied.

\begin{figure}[htp!]
 \centering
 \includegraphics[width=1.1\linewidth]{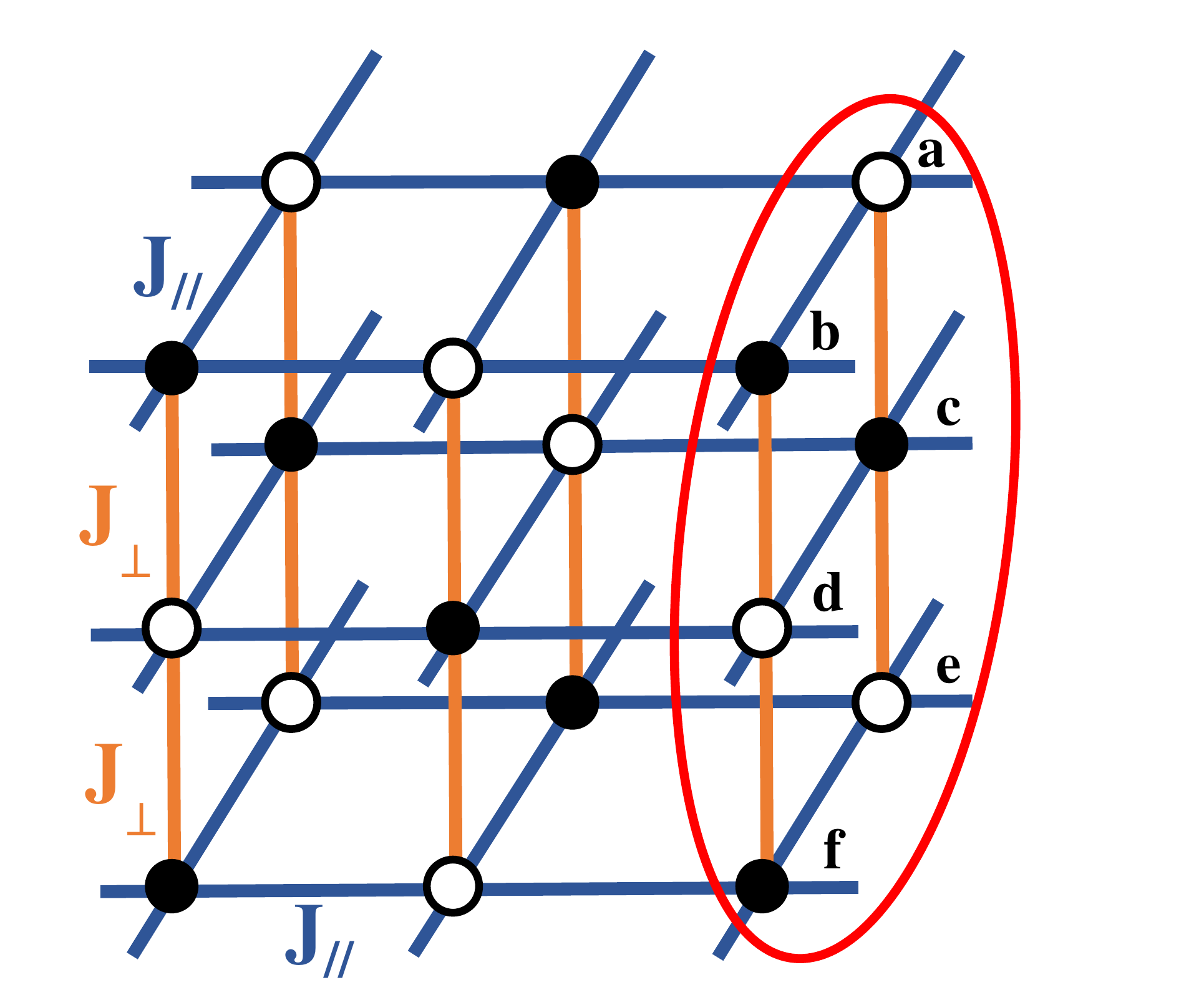}
 \caption{The trilayer antiferromagnetic Heisenberg model. Black circles represent spin up, white circles represent spin down, and all spins are $S=1/2$. In the x, y, and z directions, only nearest-neighbor interactions are considered, with in-plane coupling $J_{//}$ (blue bonds) and inter-plane coupling $J_{\bot}$ (orange vertical bonds). In the horizontal x and y directions, periodic boundary conditions are applied, while in the vertical z direction, it is an open boundary condition. The red quadrilateral represents the magnetic cell.}
 \label{model}
\end{figure}

As observed in bilayer systems, the stacking of two-dimensional layers offers additional degrees of freedom and may result in distinctive magnetic characteristics. At absolute zero, N\'eel ordered phases and magnetically disordered phases are identified, with the antiferromagnetic order-disorder transition point established at $(J_{\bot}/J_{//})_{c}=2.5220(1)$~\cite{weihong1997various,wang2006high}. Even for the more intricate dimer-diluted bilayer square lattice, by changing the interaction ratio $J_{\bot}/J_{//}$, one can detect a N\'eel ordered phase, a gapless quantum glass phase, and a gapped quantum paramagnetic phase~\cite{ma2015mott}. In addition, it has been seen that the bilayer materials $BaCuSi_{2}O_{6}$, $Cs_{3}Cr_{2}Br_{9}$ and $MnBi_{2}Te_{4}$ have a spin-singlet dimerized ground state and a singlet-to-triplet dimer excitation ~\cite{sasago1997temperature,leuenberger1984spin} that are different from those in monolayer materials. Under high pressures, ranging from $14.0$ GPa to $43.5$ GPa, the bilayer structure of $La_{3}Ni_{2}O_{7}$ single crystals has been found to exhibit superconducting characteristics~\cite{sun2023signatures,luo2023bilayer,li2019intrinsic}.

When there are three layers, the z-axis lacks symmetry, making calculations difficult. The trimer model, a simplified trilayer structure, is a good study starting point. The energy spectra of the one-dimensional antiferromagnetic trimer chain highlights two key excitation modes: the doublon mode in the intermediate-energy regime and the quarton mode in the high-energy regime~\cite{cheng2022fractional}. The same was also found in the two-dimensional trimeric system~\cite{chang2024magnon}. The ground state of a trilayer is a trivial ordered state~\cite{hu2013exchange,chatterjee2022inter,weber2022quantum}; however, its excited states demonstrate a variety of physical properties by adjusting the interaction strength between layers, including compounds $Bi_{2}Sr_{2}Ca_{2}Cu_{3}O_{10 + x}$, $YBa_{2}Cu_{3}O_{6+\epsilon }$, $HgBa_{2}Ca_{2}Cu_{3}O_{8}$ and $CrI_{3}$~\cite{roth1987structure,shamray2009crystal,wang2021systematic}. The discovery of high-temperature superconductivity in the bilayer nickelate $La_{3}Ni_{2}O_{7}$~\cite{sun2023signatures} has sparked interest in $La_{4}Ni_{3}O_{10}$, a compound with magnetic particles and a trilayer structure. $La_{4}Ni_{3}O_{10}$ can induce a trimeric lattice, with metal density waves leading to an unusual metal-to-metal transition~\cite{zhang2020intertwined,qin2024frustrated}; however, the results of its inelastic neutron scattering measurements have not yet been published.

Our research investigated the trilayer antiferromagnetic Heisenberg model shown in Fig.~\ref{model}. Both numerical simulation and theoretical analysis produced consistent findings. To account for intralayer and interlayer interactions, we set $J_{\bot}=1$ and establish a tuning parameter $g={J_{//}}/{J_{\bot}}$. The squared sublattice magnetizations indicate that the outer layer has the biggest magnetization at $g=1$, whereas the inner layer has it at $g=\frac{5}{3}$ at the thermodynamic limit. Moreover, we plotted the magnetic excitation spectra. As $g$ decreases, the excitation spectrum evolves from a single magnon spin wave to a low-energy spin wave, intermediate-energy doublon flat band ($\omega=1$), and high-energy quarton flat band ($\omega=1.5$), which become distinguishable when $g$ is reduced to 0.02. Studying trilayer magnetic excitations provides a theoretical foundation for trilayer magnetic materials and sheds light on high-temperature superconductivity.

The remaining parts follow this pattern. Section \ref{Sec:Model} outlines the trilayer Heisenberg model composition and the numerical and theoretical methodologies used for measurement. In Sec.\ref{Sec:Numerical}, we present the numerical results of the model, including the squared sublattice magnetizations and magnetic excitation spectra. In Sec.\ref{Sec:Theoretical}, we analyze theoretically calculated dispersion relations. In the final section, \ref{Sec:Summary}, we provide our study findings and tentative plans for future research.

\section{Model and Method}
\label{Sec:Model}
\subsection{Model Hamiltonian}
\label{Sec:Hamiltonian}
\begin{figure}[ht]
\centering
\includegraphics[width=1.0\linewidth]{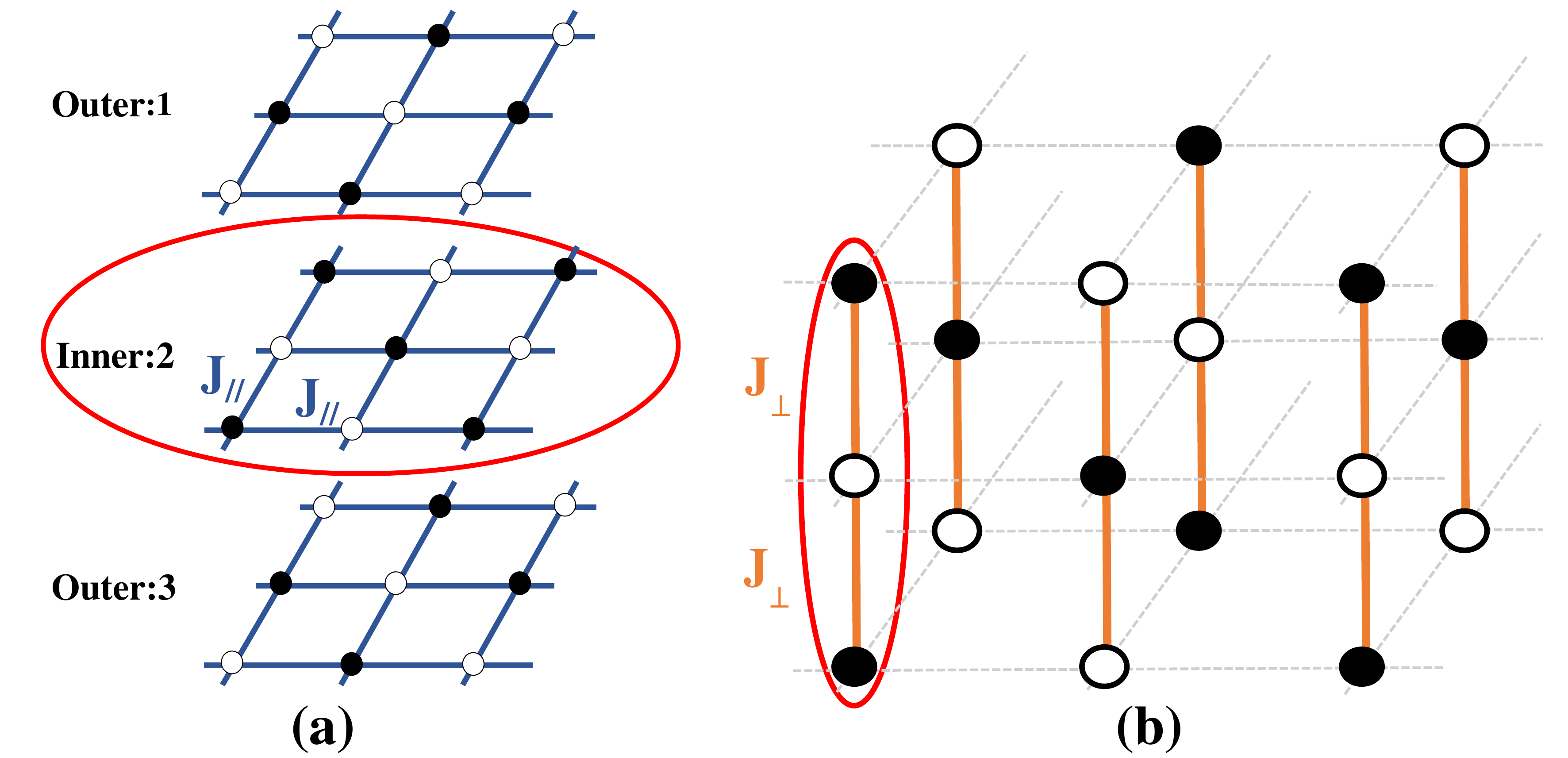}
 \caption{(a) The two-dimensional square lattice (in the red circle). The first and third layers share the same physical environment and are referred to as the outer layers, while the second layer is called the inner layer. (b) The non-contact vertical trimer structure (in the red circle). A trimer surface layer is formed when three vertical spins polymerize into a trimer.}
\label{extreme}\centering
\end{figure}

The Hamiltonian of this $S=1/2$ trilayer antiferromagnetic Heisenberg model is given by
\begin{equation}
H = J_{//}\sum\limits_{\left\langle {i,j} \right\rangle} {\mathbf{S}}_{i} \cdot {\mathbf{S}}_{j} + J_{\bot}\sum\limits_{\left\langle {i,j} \right\rangle^{\prime}} {\mathbf{S}}_{i} \cdot {\mathbf{S}}_{j},
\label{Eq:Hmlt}
\end{equation}
where $S_i$ denotes the spin operator on each site i, ${\left\langle {i,j}\right\rangle}$ denotes nearest-neighbor sites on the intralayer bonds, and ${\left\langle {i,j} \right\rangle^{\prime}}$ denotes the interlayer bonds. We use $J_{//}$ and $J_{\bot}$ to indicate intralayer and interlayer coupling strengths. This is seen when $J_{\bot}=0$ ($g=\infty $), the trilayer model degenerates into three two-dimensional (2D) square lattices that have been studied in great detail, as shown in Fig.~\ref{extreme} (a). In contrast, when $J_{//}=0$ ($g=0$), only $J_{\bot}$ in the vertical direction remains, resulting in non-contact trimers in the horizontal direction, as shown in Fig.~\ref{extreme} (b). Trimers may now be considered a single, bigger spin, changing the model into a renormalized two-dimensional plane with $S=1/2$.

\subsection{Quantum Monte Carlo}
\label {Sec:QMC}

The stochastic series expansion quantum Monte Carlo (SSE-QMC) approach makes physical quantity measurements easier and avoids systematic mistakes that might lower accuracy. The sign problem does not apply to our model since it only considers the closest neighbor interactions.

SSE-QMC works by Taylor expanding the partition function, as shown in the formula: 
\begin{equation}
Z=Tr\left \{ e^{-\beta H } \right \} =\sum_{\alpha }\sum_{n=0}^{\infty }\frac{\left ( \beta \right )^{n}}{n!}\left \langle\alpha \left | \left ( -H \right )^{n} \right | \alpha \right \rangle. 
\label{Eq:Z}
\end{equation}
After several updates, the system reaches equilibrium and is then amenable to representation by operators for physical observations like ground-state energy and spin correlation.

In addition to thermodynamic quantities, SSE-QMC can calculate the dynamic spin structure factor. SSE-QMC sampling is used to obtain imaginary-time correlation functions. These functions are defined by:
\begin{equation}
G_{q}(\tau )=3\left \langle S_{-q}^{z}(\tau ) S_{q}^{z}(0) \right \rangle.
\label{Eq:G}
\end{equation} 
The spin operators $S^{x}$, $S^{y}$, and $S^{z}$ are equivalent in the Hamiltonian, resulting in a factor of 3 in it. Since the trilayer model lacks periodic boundary conditions in the z-direction, Fourier expansion cannot be applied in this direction. We must compute $G_{q}(\tau)$ for each layer separately, then sum the layer values to get the $G_{q}(\tau)$ of the overall model. 

The term $S_{q}^{z}$ represents the Fourier transform of the z-component spin operator for each layer, expressed as \begin{equation}
S_{q}^{z}=\frac{1}{\sqrt{N}} \sum_{i=1}^{N} e^{-i \vec{r}_{i} \cdot \vec{q}} S_{i}^{z}. 
\label{Eq:sqz}
\end{equation} 

We use the method of stochastic analytic continuation to reconstruct the spectral function $S\left (q,\omega\right )$ from a sequence of imaginary-time points for $G_{q}(\tau )$:
\begin{equation}
G_{q}(\tau)=\frac{1}{\pi} \int_{-\infty}^{\infty} d \omega S(q, \omega) e^{-\tau \omega}. 
\label{Eq:Gq}
\end{equation} 
We parametrize the spectrum using $N_{\omega}$ delta functions in the continuum:
\begin{equation}
S(q, \omega)=\sum^{N{\omega } }_{i=0} a_{i}\delta (\omega -\omega_{i}).
\label{Eq:Sqw}
\end{equation}

\subsection{Linear Spin Wave Theory}
\label{Sec:LSWT}
The dispersion relations of the trilayer antiferromagnetic Heisenberg model are calculated using the linear spin-wave theory (LSWT).

To ensure reproducibility, we chose six lattices to construct a macromolecular cell (red circle in Fig.~\ref{model}). 

For this model, the Holstein-Primakoff boson method can be used to calculate the dispersion relation~\cite{yao2008magnetic}. We quantize the above Hamiltonian using Holstein-Primakoff bosons in the momentum space and disregard the higher-order components to achieve the low-energy approximations:
\begin{equation}
H=E_{Cl}+\sum_{k,\left\langle {i,j} \right\rangle} C_{ii}a_{k,i}^{\dagger}a_{k,i}+C_{ij}(a_{k,i}a_{k,j}+a_{k,i}^{\dagger}a_{k,j}^{\dagger}),
\label{Eq:H'}
\end{equation}
where $E_{Cl}=(-12J_{//}-4J_{\bot})NS^{2}$ is the ground state energy and $i,j$ are spin indexes in the unit cell.
Since the Hamiltonian contains off-diagonal terms, we diagonalize it using the extended Bogoliubov transformation:
\begin{equation}
b_{k,i}=\sum_{j} m_{ij}a_{k,j}+m'_{ij}a_{k,j}^{\dagger}.
\label{Bogoliubov}
\end{equation}
Only one quasiparticle may result from the transformation due to the symmetry of the Hamiltonian:
\begin{equation}
b_{k,i}=m_{11}a_{k1}+m_{12}a_{k2}^{\dagger}+m_{13}a_{k3}+m_{14}a_{k4}^{\dagger}+m_{15}a_{k5}+m_{16}a_{k6}^{\dagger}.
\label{Eq:a'}
\end{equation}

The transformation coefficients $m_{ki}$ can be obtained by setting the determinant zero from the equation of motion ~\cite{tan2023spin,carlson2004spin,yao2008magnetic}
\begin{equation}
i\hbar \dot{b}_{k,i}=[b_{k,i}, H]. 
\label{Eq:qua}
\end{equation}

The diagonalized Hamiltonian is
\begin{equation}
H=E_{Cl}+E_{0}+\sum_{k,i}\omega_i(k)b_{k,i}^{\dagger}b_{k,i},
\label{Hfinal}
\end{equation}
where $\omega_i(k)$ is the spin wave dispersion, and $E_{0}$ is the quantum zero-point energy correction.

\section{Numerical Results}
\label{Sec:Numerical}
In our QMC-SAC calculations, each layer of the trilayer Heisenberg model has dimensions $L=L_x = L_y$, with both $L_x$ and $L_y$ being even numbers, resulting in a total size of $N=3 L^{2} $. After rigorous convergence testing, we set the inverse temperature to $\beta = 6 L$, and each data set is equipped with 500 bins to reduce systematic error. These measures ensure the precision and reliability of the numerical calculations.

\subsection{Sublattice Magnetizations}
\label {Sec:Sublattice Magnetization}
In this subsection, we present the squared sublattice magnetizations of $J_{\bot}$ in different cases of the trilayer model, along with their extrapolated results.

The squared sublattice magnetizations $m_{s}^{2}$ are defined as
\begin{equation}
m_{s}^{2}=\frac{3}{N^{\prime^{2}}}\left \langle \left ( \sum_{i=1}^{N^{\prime}}\phi _{i} S_{i}^{z} \right )^{2} \right \rangle, 
\label{Eq:ms2}
\end{equation}
where $\phi _{i}=\pm 1$ signifies staggered phase factors, while the factor of $3$ is attributed to the isotropic strength encompassing all three components of the spin. For the complete model, $N^{\prime}=3 L^{2}$, while for a single layer, $N^{\prime}= L^{2}$.

\begin{figure}[ht]
\centering
 \includegraphics[width=1\linewidth]{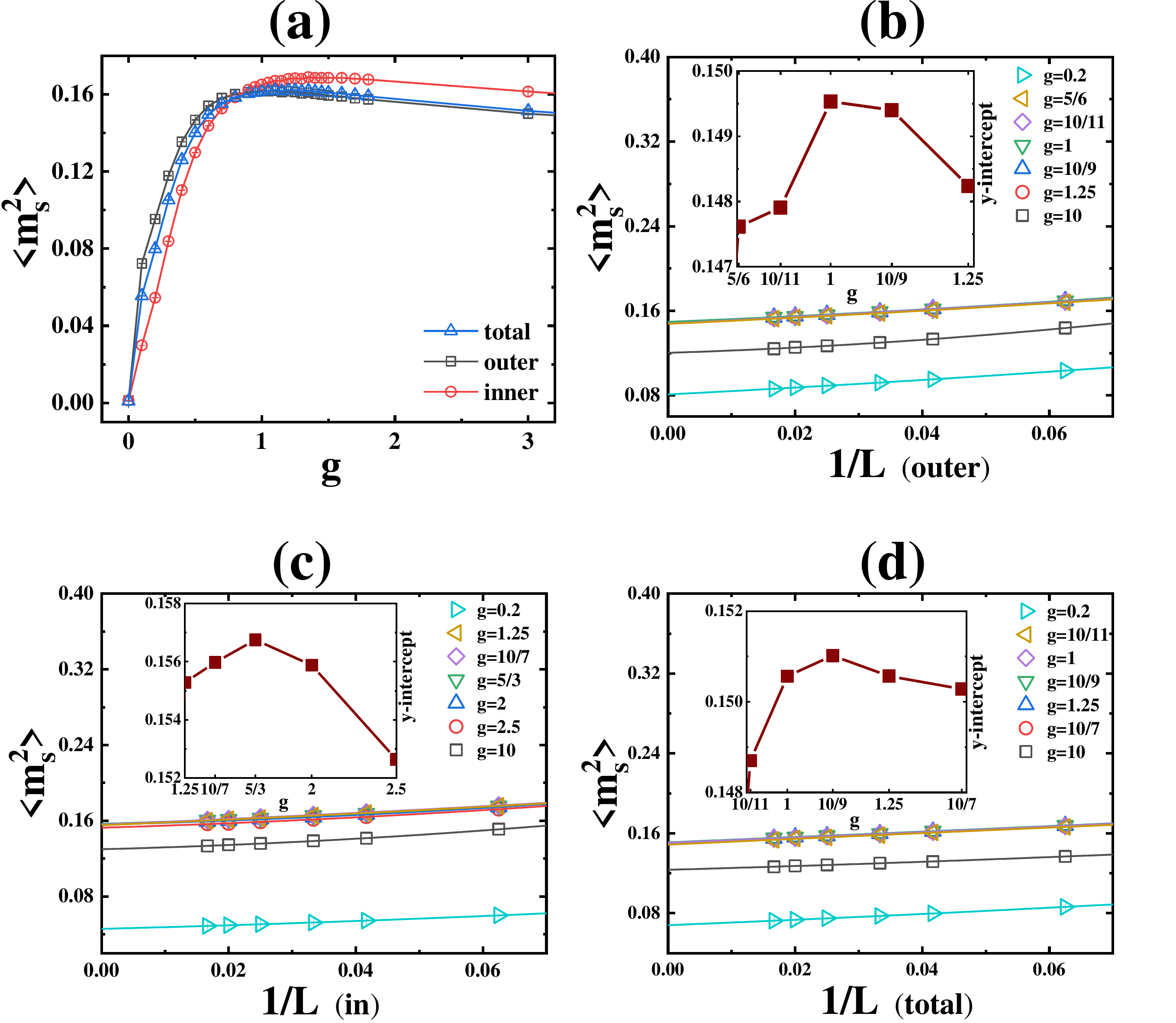}
 \caption{(a) The squared sublattice magnetizations vary with $g$ in the outer layer, inner layer, and the entire trilayer model. (b), (c), and (d) show the extrapolation of the squared sublattice magnetizations as a function of $1/L$ for different $g$ conditions. (b) corresponds to the outer layer, (c) corresponds to the inner layer, and (d) corresponds to the average total for three layers. The inset graphs show $g$ corresponding to the maximum of the intercept. In the above four graphs, the error bars are smaller than the size of the symbols.}
\label{fig:ms2}
\end{figure}

Figure~\ref{fig:ms2} shows the squared sublattice magnetizations versus $g$ (with $J_{\bot}=1$) for a system size of $L=24$. For any layer, $m_{s}^{2}$ first rises and then falls as $g$ rises. After fitting three sets of data using a second-order polynomial function, we discovered that the inner maximum is around $g=\frac{5}{3}$ and the outer maximum is about $g=1$. Additionally, the maximum of the trilayer is centered at $g=\frac{10}{9}$. Improved intralayer antiferromagnetic interactions enhance long-range antiferromagnetic order, with $g$ increasing from 0 to roughly 1. The model shows better alignment in the three directions when $g$ approaches 1, signifying the highest magnetic order. As the parameter $g$ increases sufficiently, the model evolves into three almost isolated two-dimensional square lattices. The reduction of dimensions leads to a decrease in long-range antiferromagnetism. The magnetizations of the outer layers are higher than those of the inner layer when $g$ is between 0 and 1 (the black line is higher than the red line). However, the conclusion is reversed from 1 to 5. The inner layer spins have a coordination number of 6, with 4 horizontal and 2 vertical neighbors. In contrast, the outer layer spins have a coordination number of 5, with 4 horizontal and 1 vertical neighbors. As $g$ increases, the inner layer spins are more significantly impacted.

To precisely determine the $g$ corresponding to the extremum points for the three situations, we use finite-size extrapolation for sizes L = 16, 24, 30, 40, 50, and 60. The y-axis intercepts on Figs.\ref{fig:ms2} (b), (c), and (d) show the thermodynamic limit values derived from second-order polynomial fittings. These match the pattern shown in Fig.~\ref{fig:ms2} (a): the maximum magnetization in the outer layer ($m_{s}^{2}=0.1494(4)$) is observed at $g=1$, while the maximum magnetization in the inner layer ($m_{s}^{2}=0.1567(5)$) occurs at $g=\frac{5}{3}$, and the entire trilayer model reaches its maximum magnetization ($m_{s}^{2}=0.1510(2)$) when $g=\frac{10}{9}$. As the model approaches the trimerization structure ($g=0.2$, light blue line), the squared sublattice magnetization of the outer layers ($m_{s}^{2}=0.0809(3)$) in the thermodynamic limit exceeds that of the inner layer ($m_{s}^{2}=0.0457(2)$). As $g$ decreases to 0, the model gradually evolves into a vertical trimerization structure. Under these conditions, the inner layer spins pair with the coordinated spins of either the upper or lower layer to form a singlet, resulting in an unpaired spin in the outer layer. The unpaired spins yield greater magnetization in the outer sublattice.

The squared sublattice magnetization primarily reflects the ground-state properties of the model. To conduct a more comprehensive study of the model, we further analyze its excited states by calculating the dynamic structure factor. The results indicate that the maximum of the squared sublattice magnetization occurs at $g\approx1$ (see Fig.~\ref{fig:ms2} (a)), which also gives more stable spin wave excitations, as shown in Fig.~\ref{fig:sac} (b).

\subsection{Magnetic Excitations}
\label {Sec:Magnetic Excitation}

\begin{figure}[ht]
 \centering
 \includegraphics[width=0.5\linewidth]{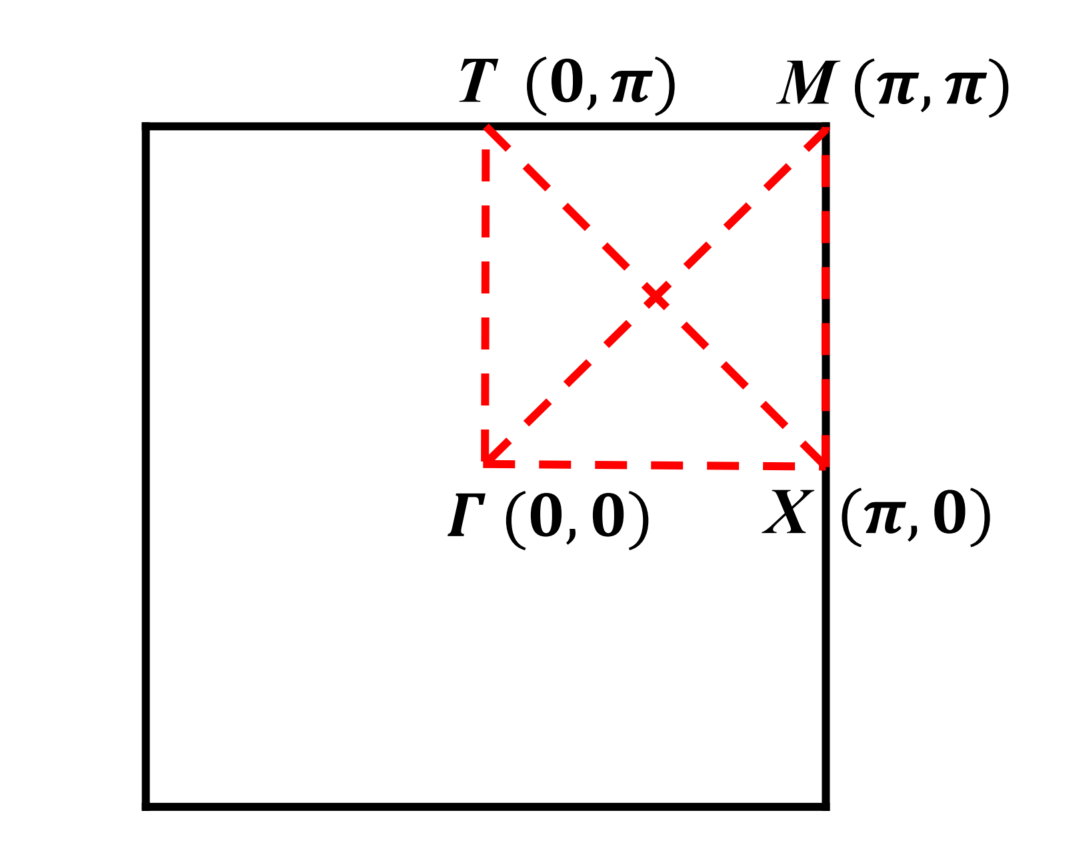}
 \caption{The full Brillouin zone (BZ) is given, and the highly symmetrical path (${\Gamma}$-X-M-${\Gamma}$-T-X) selected in the calculation is demarcated.}
 \label{fig:BZ}
\end{figure}

\begin{figure*}
\centering
\includegraphics[width=1\textwidth]{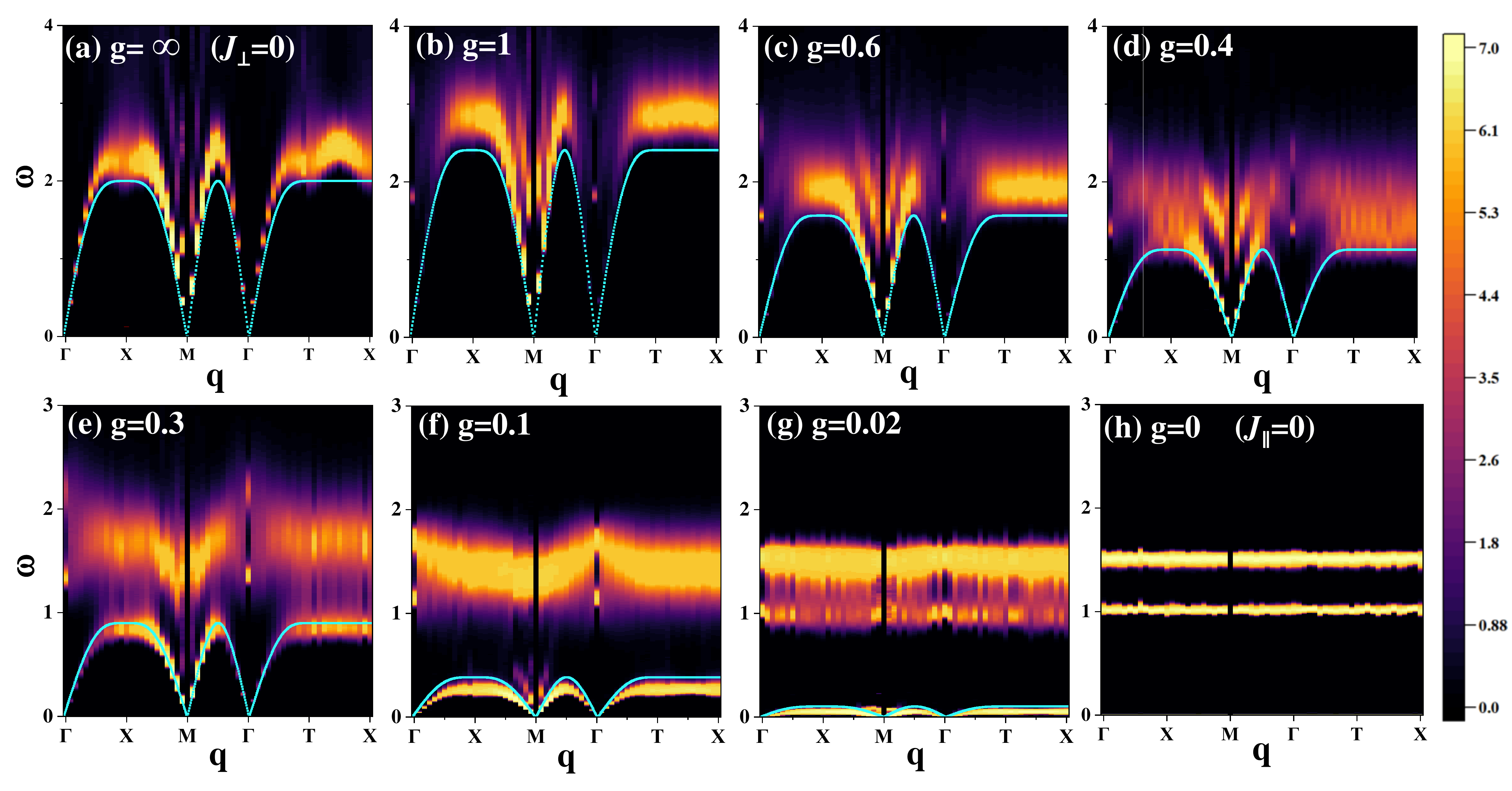}
\caption{The dynamic spin structure factor $S(q,\omega)$ is derived from QMC-SAC calculations for the trilayer model with varying values of $g$. The figures are arranged from left to right for $g$ values ranging from infinity to zero. Figures (a)-(d) have a vertical scale of $0-4$, and figures (e)-(h) have a vertical scale of $0-3$.The blue lines within the graph represent the results derived from the linear spin wave theory [as given by Eq.(\ref{result})], and these theoretical results are compared with the outcomes of numerical calculations.}
\label{fig:sac} 
\end{figure*}

In this subsection, we present the magnetic excitation spectra of the trilayer antiferromagnetic Heisenberg model via QMC-SAC calculations. We set system size $L=24$, inverse temperature $\beta=6 L=144$. To visually represent the $S\left (q,\omega\right)$, we follow a high-symmetry trajectory in the Brillouin zone: $\Gamma (0,0)\to X(\pi ,0)\to M(\pi ,\pi )\to \Gamma (0 ,0)\to T (0 ,\pi )\to X(\pi ,0)$. The above path is depicted in Fig.~\ref{fig:BZ}. Figure~\ref{fig:sac} demonstrates how the dynamic spin structure factor $S\left (q,\omega\right )$ varies with changes in $g$. We choose $g$ at $\infty $ ($J_{\bot}=0$), 1, 0.6, 0.4, 0.3, 0.1, 0.02, and 0 to illustrate the advancement. Due to the difference of three orders of magnitude between the minimum and maximum of $S(q, \omega)$, a piecewise function was applied to clearly see both areas of high and low intensity simultaneously in Fig.~\ref{fig:sac}. The low-intensity region has a linear distribution for values below $U_0$. A logarithmic scale is applied when the value surpasses this threshold, as denoted by $U = U_0 + \log_{10}(S(q, \omega)) - \log_{10}(U_0)$. After multiple attempts, we choose $U_0 = 6$. This outcome can be compared to the neutron scattering excitation spectra of material with a trilayer structure to confirm the accuracy of the theoretical results.

First, we must verify the findings of $g=J_{//}/J_{\bot}=\infty$ as presented in Fig.~\ref{fig:sac} (a). The trilayer Heisenberg model separates into three unconnected single-layer two-dimensional square lattices, as illustrated in Fig.~\ref{extreme} (a). The dispersion relation, indicative of the expansion of linear spin waves~\cite{yao2008magnetic}, has been determined by previous researchers and is expressed as follows: 
\begin{equation}
\omega_{AF} =2 J_{//} S \sqrt{4-(\cos k_{x} + \cos k_{y} )^{2}.} 
\label{Eq:wAF}
\end{equation}
This equation aligns with the acoustic magnon band in our result for LSWT, as shown by the blue line in Fig.~\ref{fig:sac} (a). The spectra meet with the established pattern of linear spin waves, where the points $\Gamma$ and $M$ show gapless modes and the band top approaches $\omega=2$. The LSWT approximation causes band top matching differences. The LSWT cannot describe the high-energy spectra obtained by QMC-SAC. Since LSWT ignores several high-order factors throughout the calculation process, the theoretical results are slightly lower than actual results in the high-energy excitation region ~\cite{shao2017nearly,chang2024magnon}. The spectra display a gapless Goldstone mode at the point $M(\pi,\pi)$, distinguished by maximized spectral weight shown in a bright yellow color and indicating divergence. The uneven and arched excitation from $T (0 ,\pi )$ to $X(\pi ,0)$ is due to the decay of spin waves into other excitations at T and X, which suppresses the high-energy excitations at these two points. This has been analyzed in many previous results~\cite{dalla2015fractional,headings2010anomalous}. All findings fit with the SAC spectra observed in previous research ~\cite{shao2017nearly}. It is important to acknowledge that the gapless point at $M(\pi,\pi)$ requires a very large $\beta$ to achieve convergence results. Consequently, we did not include them in Fig.~\ref{fig:sac}. Nevertheless, we can infer their behavior from the points surrounding it.

Next, from Fig.~\ref{fig:sac} (a) to (b), $J_{\bot}$ decreases from $\infty$ to 1. The number of internal interactions in the model rises dramatically throughout this process, and its structure moves from a two-dimensional plane to a quasi-two-dimensional one. This change results in a higher excitation energy and an expansion of the bandwidth. As $g$ continues to decrease, the band top systematically declines, but the spectral weight increases. This singular continuous spectrum begins to split into a high-energy part and a low-energy branch as $g\approx0.3$. The dispersion relation derived from the linear spin wave method (shown by the blue line in Fig.~\ref{fig:sac}) accurately corresponds to the low-energy spin wave spectrum in the bottom part. The high-energy continuous spectrum reflects the characteristics of the internal coupling within the trimer. The spectral weight is directly proportional to $J_{\bot}$, resulting in a gradual increase, while the broadening concurrently reduces. This evidence indicates that high-energy quantum excitations can only be significantly observed when the value of $J_{\bot}$ is comparatively large. When $g\approx0.02$, the high-energy part clearly splits into two flat bands, and three distinctly separated spectral lines are seen in Fig.~\ref{fig:sac} (g). The characteristic spectrum around $\omega \approx 1$, associated with the quasiparticle, is termed the doublon, while the spectrum around $\omega \approx 1.5$, corresponding to another type of quasiparticle, is called the quarton~\cite{cheng2022fractional}.

Finally, let us discuss another extreme case where $g=J_{//}/J_{\bot}=0$, and only the vertical coupling strength within the unit cell $J_{\bot}$ remains. Currently, the model consists of isolated vertical trimers, each can be considered as an effective spin of $S=1/2$, as illustrated in Fig.~\ref{extreme} (b). The excitations within the trimers are localized and without any magnons. In Fig.~\ref{fig:sac} (h), the classical spin wave excitations disappear. Near $\omega \approx 1$ and $\omega \approx 1.5$, two continuous spectral bands exhibit minimal energy broadening and significant power weight. These bands represent the internal excitation states of the trimer. These two characteristic spectral lines perfectly match previous studies~\cite{cheng2022fractional}, not only demonstrating the accuracy of our SAC calculations but also elucidating the similarity between trimer chains and isolated vertical trimers. The ground state of a trimer is a doublet with energy $E_{0}=-J_{\bot}$, whereas the first and second excited states are a doublet with energy $E_{1}=0$ and a quartet with energy $E_{2}=J_{\bot}/2$, respectively. The two low-energy doublets include only singlets and unpaired spins, while the higher-energy quadruplet excitations include either a triplet pair with an unpaired spin or three unpaired spins collectively.

\section{Analysis}
\label{Sec:Theoretical}

As mentioned earlier, we have selected six lattice points as a macromolecular cell, so we can obtain three dispersion relationships: the acoustic mode ($\omega_{1}$), the lower energy optical mode ($\omega_{2}$), and the higher energy optical mode ($\omega_{3}$). The spin wave dispersion $\omega$ is given by 
\begin{equation}
  \begin{cases}
   \omega_{1}=S\sqrt{A_{k}-B_{k}} \\
   \omega_{2}=S\sqrt{(4J_{//}+J_{\bot})^{2}-(2J_{//}\cos k_{x}+2J_{//}\cos k_{y})^{2}} \\ 
   \omega_{3}=S\sqrt{A_{k}+B_{k}},
  \end{cases}
\label{result}
\end{equation}
where
\begin{equation}
  \begin{aligned}
   &A_{k}=(4J_{//})^{2}+\frac{J_{\bot}^{2}}{2}+12J_{//}J_{\bot}-4J_{//}^{2}(\cos k_{x}+\cos k_{y})^{2} \\
   &B_{k}=4J_{\bot}\sqrt{2J_{//}^{2}(\cos k_{x}+\cos k_{y})^{2}+J_{//}^{2}+\frac{J_{\bot}^{2}}{64}+\frac{3}{4}J_{//}J_{\bot}}.
  \end{aligned}
\label{AkBk}
\end{equation}
The final results indicate that the model shows a certain degree of decoupling, with the inner and outer layers exhibiting identical dispersion relations. The three dispersion relationships represent acoustic and optical modes. The acoustic mode depicts the movement of cell centroids, whereas the optical mode describes the relative motion between spins within the cell. 

\begin{figure}[ht]
 \centering
 \includegraphics[scale=0.35]{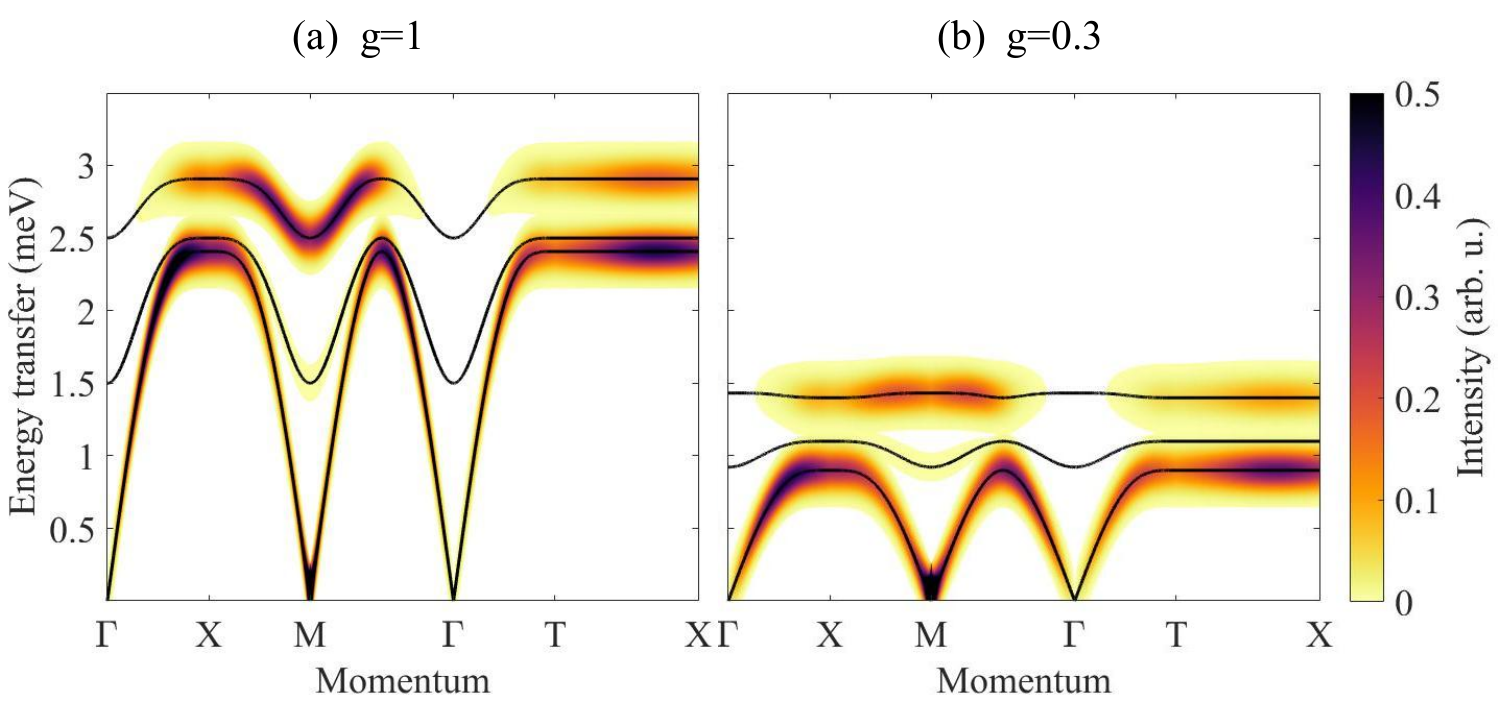}
 \caption{The solid lines in the figure illustrate three dispersion curves at different $g$ values: (a) $g=1$, partial separation; (b) $g=0.3$, fully separated. The spectral plot represents the simulation results of SpinW.}
 \label{fig:ek}
\end{figure}

In Fig.~\ref{fig:ek}, the theoretically derived dispersion relations show excellent agreement with the SpinW simulation results. In the limit of $g\to \infty$ $(g=J_{//}/J_{\bot})$, the three dispersion relations converge to those described by Eq.\ref{Eq:wAF} of the two-dimensional square lattice. As $g$ decreases towards 1, the model becomes more three-dimensional. This change is accompanied by an increase in the influence of interlayer coupling, leading to an overall increase in excitation energies. At $g=1$, the highest energy mode ($\omega_{3}$) separates significantly from the other two lower-energy modes. As shown in Fig.~\ref{fig:ek} (a), a clear separation between the high-energy and low-energy regions is observed. Fig.~\ref{fig:ek} (b) illustrates that at $g=0.3$, a distinction emerges between the optical and acoustic modes. Notably, the spin wave dispersion becomes a flat band when $g=0$.

Theoretical acoustic modes ($\omega_{1}$) exhibit consistency with the numerical results, but the optical modes do not match the high-energy portion of the numerical calculations. Since the calculation only considers the lowest-order terms in the Hamiltonian, the resulting dispersion is a low-energy approximation, accurately describing the spin excitation spectra only in the low-energy regime.

The associated spin-wave velocities are
\begin{equation}
\begin{aligned}
&v_{x}=v_{y}=\\ 
&\sqrt{48J_{//}^{2} +2J_{\bot}^{2}+48J_{//}J_{\bot}-2J_{\bot}\sqrt{192J_{//}^{2} +48J_{//}J_{\bot}+J_{\bot}^{2}}}\\.
\end{aligned}
\label{vx}
\end{equation}
The spin-wave velocity exhibits a linear dependence on $g$. Consequently, as $J_{//}$ increases, the wave velocity increases, and as $J_{\bot}$ increases, it decreases. These results are consistent with the theoretical prediction that strong $J_{\bot}$ favors the formation of trimer bound states, whereas strong $J_{//}$ facilitates spin-wave propagation within the horizontal plane. 

A key issue of the linear spin wave technique is to use operator characteristics to turn the Schr$\ddot{o}$dinger equation into a linear system. Compared to other approaches, the linear spin wave method is easier to compute, simpler to program, and yields an exact analytical answer. However, it can only calculate a few physical quantities and is suitable only for low-energy approximations.

\section{Summary and Discussion}
\label{Sec:Summary}

In this study, we investigated magnetic excitations of the trilayer antiferromagnetic Heisenberg model (Fig.~\ref{model}). We computed the squared sublattice magnetizations of each layer and found that the value of $g$ corresponding to the maximum magnetization of the outer layer is smaller than that of the inner layer. We employed finite-size extrapolation to verify this discovery. Subsequently, we calculated the dynamic spin structure factor of the whole model using SSE-QMC and SAC methods to investigate the magnetic excitation spectra. We observed that the spin wave excitation spectrum separates into low-energy and high-energy sections at $g\approx0.3$ as the parameter $g$ decreases. The high-energy component further splits into doublon and quarton branches when $g$ is small. To validate our numerical results, we derived the spin wave dispersions of the trilayer Heisenberg model from the LSWT. The resulting dispersion relations agree well with the SpinW simulations, exhibiting one low-energy acoustic branch and two high-energy optical branches. In the model, when $g$ is sufficiently large, the LSWT results match well with the SSE-QMC and SAC calculations. However, as $g$ approaches 0, doublon and quarton excitations become increasingly dominant, resulting in a worse alignment between them. Notably, the low-energy acoustic branch mainly coincides with the low-energy spin wave component of the excitation spectra, as predicted by the semiclassical spin-wave theory. 

Going forward, our theoretical conclusions may be confirmed by inelastic neutron scattering measurements on $Bi_{2}Sr_{2}Ca_{2}Cu_{3}O_{10 + x}$, $HgBa_{2}Ca_{2}Cu_{3}O_{8}$, and $La_{4}Ni_{3}O_{10}$~\cite{abbas2012effect,ali2024study}. In other trilayer materials, quantum excitations linked with spin waves should be observable when vertical interactions are substantial. Additionally, ultracold atoms in optical lattices, Rydberg atom systems, and quantum networks offer exciting options for researching the trilayer model beyond the current limits~\cite{adams2019rydberg,wirth2011evidence,gross2017quantum}.

\begin{acknowledgments}
We thank Han-Qing Wu, Jun-Qing Cheng, Xuyang Liang, Muwei Wu and Jiayuan Zhang for their helpful discussions. This project is supported by NKRDPC2022YFA1402802, NSFC-12494591, NSFC-92165204, Leading Talent Program of Guangdong Special Projects (201626003), Guangdong Provincial Key Laboratory of Magnetoelectric Physics and Devices (No. 2022B1212010008), Research Center for Magnetoelectric Physics of Guangdong Province (2024B0303390001), and Guangdong Provincial Quantum Science Strategic Initiative (GDZX2401010).
\end{acknowledgments}

\bibliography{main}

\begin{thebibliography}{42}%
\makeatletter
\providecommand \@ifxundefined [1]{%
 \@ifx{#1\undefined}
}%
\providecommand \@ifnum [1]{%
 \ifnum #1\expandafter \@firstoftwo
 \else \expandafter \@secondoftwo
 \fi
}%
\providecommand \@ifx [1]{%
 \ifx #1\expandafter \@firstoftwo
 \else \expandafter \@secondoftwo
 \fi
}%
\providecommand \natexlab [1]{#1}%
\providecommand \enquote  [1]{``#1''}%
\providecommand \bibnamefont  [1]{#1}%
\providecommand \bibfnamefont [1]{#1}%
\providecommand \citenamefont [1]{#1}%
\providecommand \href@noop [0]{\@secondoftwo}%
\providecommand \href [0]{\begingroup \@sanitize@url \@href}%
\providecommand \@href[1]{\@@startlink{#1}\@@href}%
\providecommand \@@href[1]{\endgroup#1\@@endlink}%
\providecommand \@sanitize@url [0]{\catcode `\\12\catcode `\$12\catcode `\&12\catcode `\#12\catcode `\^12\catcode `\_12\catcode `\%12\relax}%
\providecommand \@@startlink[1]{}%
\providecommand \@@endlink[0]{}%
\providecommand \url  [0]{\begingroup\@sanitize@url \@url }%
\providecommand \@url [1]{\endgroup\@href {#1}{\urlprefix }}%
\providecommand \urlprefix  [0]{URL }%
\providecommand \Eprint [0]{\href }%
\providecommand \doibase [0]{https://doi.org/}%
\providecommand \selectlanguage [0]{\@gobble}%
\providecommand \bibinfo  [0]{\@secondoftwo}%
\providecommand \bibfield  [0]{\@secondoftwo}%
\providecommand \translation [1]{[#1]}%
\providecommand \BibitemOpen [0]{}%
\providecommand \bibitemStop [0]{}%
\providecommand \bibitemNoStop [0]{.\EOS\space}%
\providecommand \EOS [0]{\spacefactor3000\relax}%
\providecommand \BibitemShut  [1]{\csname bibitem#1\endcsname}%
\let\auto@bib@innerbib\@empty
\bibitem [{\citenamefont {Mermin}\ and\ \citenamefont {Wagner}(1966)}]{1mermin1966absence}%
  \BibitemOpen
  \bibfield  {author} {\bibinfo {author} {\bibfnamefont {N.~D.}\ \bibnamefont {Mermin}}\ and\ \bibinfo {author} {\bibfnamefont {H.}~\bibnamefont {Wagner}},\ }\bibfield  {title} {\bibinfo {title} {Absence of ferromagnetism or antiferromagnetism in one-or two-dimensional isotropic heisenberg models},\ }\href@noop {} {\bibfield  {journal} {\bibinfo  {journal} {Physical Review Letters}\ }\textbf {\bibinfo {volume} {17}},\ \bibinfo {pages} {1133} (\bibinfo {year} {1966})}\BibitemShut {NoStop}%
\bibitem [{\citenamefont {Deiseroth}\ \emph {et~al.}(2006)\citenamefont {Deiseroth}, \citenamefont {Aleksandrov}, \citenamefont {Reiner}, \citenamefont {Kienle},\ and\ \citenamefont {Kremer}}]{1deiseroth2006fe3gete2}%
  \BibitemOpen
  \bibfield  {author} {\bibinfo {author} {\bibfnamefont {H.-J.}\ \bibnamefont {Deiseroth}}, \bibinfo {author} {\bibfnamefont {K.}~\bibnamefont {Aleksandrov}}, \bibinfo {author} {\bibfnamefont {C.}~\bibnamefont {Reiner}}, \bibinfo {author} {\bibfnamefont {L.}~\bibnamefont {Kienle}},\ and\ \bibinfo {author} {\bibfnamefont {R.~K.}\ \bibnamefont {Kremer}},\ }\bibfield  {title} {\bibinfo {title} {{Fe$_{3}$GeTe$_{2}$ and Ni$_{3}$GeTe$_{2}$--two new layered transition-metal compounds: crystal structures, HRTEM investigations, and magnetic and electrical properties}},\ }\href@noop {} {\bibfield  {journal} {\bibinfo  {journal} {European Journal of Inorganic Chemistry}\ }\textbf {\bibinfo {volume} {2006}},\ \bibinfo {pages} {1561} (\bibinfo {year} {2006})}\BibitemShut {NoStop}%
\bibitem [{\citenamefont {Lu}\ \emph {et~al.}(2022)\citenamefont {Lu}, \citenamefont {Yuan}, \citenamefont {Zhang}, \citenamefont {Li}, \citenamefont {Luo},\ and\ \citenamefont {Li}}]{1lu2022observation}%
  \BibitemOpen
  \bibfield  {author} {\bibinfo {author} {\bibfnamefont {F.}~\bibnamefont {Lu}}, \bibinfo {author} {\bibfnamefont {L.}~\bibnamefont {Yuan}}, \bibinfo {author} {\bibfnamefont {J.}~\bibnamefont {Zhang}}, \bibinfo {author} {\bibfnamefont {B.}~\bibnamefont {Li}}, \bibinfo {author} {\bibfnamefont {Y.}~\bibnamefont {Luo}},\ and\ \bibinfo {author} {\bibfnamefont {Y.}~\bibnamefont {Li}},\ }\bibfield  {title} {\bibinfo {title} {{The observation of quantum fluctuations in a kagome Heisenberg antiferromagnet}},\ }\href@noop {} {\bibfield  {journal} {\bibinfo  {journal} {Communications Physics}\ }\textbf {\bibinfo {volume} {5}},\ \bibinfo {pages} {272} (\bibinfo {year} {2022})}\BibitemShut {NoStop}%
\bibitem [{\citenamefont {Kim}\ \emph {et~al.}(2019)\citenamefont {Kim}, \citenamefont {Lim}, \citenamefont {Lee}, \citenamefont {Lee}, \citenamefont {Kim}, \citenamefont {Park}, \citenamefont {Jeon}, \citenamefont {Park}, \citenamefont {Park},\ and\ \citenamefont {Cheong}}]{2kim2019suppression}%
  \BibitemOpen
  \bibfield  {author} {\bibinfo {author} {\bibfnamefont {K.}~\bibnamefont {Kim}}, \bibinfo {author} {\bibfnamefont {S.~Y.}\ \bibnamefont {Lim}}, \bibinfo {author} {\bibfnamefont {J.-U.}\ \bibnamefont {Lee}}, \bibinfo {author} {\bibfnamefont {S.}~\bibnamefont {Lee}}, \bibinfo {author} {\bibfnamefont {T.~Y.}\ \bibnamefont {Kim}}, \bibinfo {author} {\bibfnamefont {K.}~\bibnamefont {Park}}, \bibinfo {author} {\bibfnamefont {G.~S.}\ \bibnamefont {Jeon}}, \bibinfo {author} {\bibfnamefont {C.-H.}\ \bibnamefont {Park}}, \bibinfo {author} {\bibfnamefont {J.-G.}\ \bibnamefont {Park}},\ and\ \bibinfo {author} {\bibfnamefont {H.}~\bibnamefont {Cheong}},\ }\bibfield  {title} {\bibinfo {title} {{Suppression of magnetic ordering in XXZ-type antiferromagnetic monolayer NiPS$_{3}$}},\ }\href@noop {} {\bibfield  {journal} {\bibinfo  {journal} {Nature communications}\ }\textbf {\bibinfo {volume} {10}},\ \bibinfo {pages} {345} (\bibinfo {year} {2019})}\BibitemShut {NoStop}%
\bibitem [{\citenamefont {Lee}\ \emph {et~al.}(2021)\citenamefont {Lee}, \citenamefont {Dismukes}, \citenamefont {Telford}, \citenamefont {Wiscons}, \citenamefont {Wang}, \citenamefont {Xu}, \citenamefont {Nuckolls}, \citenamefont {Dean}, \citenamefont {Roy},\ and\ \citenamefont {Zhu}}]{2lee2021magnetic}%
  \BibitemOpen
  \bibfield  {author} {\bibinfo {author} {\bibfnamefont {K.}~\bibnamefont {Lee}}, \bibinfo {author} {\bibfnamefont {A.~H.}\ \bibnamefont {Dismukes}}, \bibinfo {author} {\bibfnamefont {E.~J.}\ \bibnamefont {Telford}}, \bibinfo {author} {\bibfnamefont {R.~A.}\ \bibnamefont {Wiscons}}, \bibinfo {author} {\bibfnamefont {J.}~\bibnamefont {Wang}}, \bibinfo {author} {\bibfnamefont {X.}~\bibnamefont {Xu}}, \bibinfo {author} {\bibfnamefont {C.}~\bibnamefont {Nuckolls}}, \bibinfo {author} {\bibfnamefont {C.~R.}\ \bibnamefont {Dean}}, \bibinfo {author} {\bibfnamefont {X.}~\bibnamefont {Roy}},\ and\ \bibinfo {author} {\bibfnamefont {X.}~\bibnamefont {Zhu}},\ }\bibfield  {title} {\bibinfo {title} {{Magnetic order and symmetry in the 2D semiconductor CrSBr}},\ }\href@noop {} {\bibfield  {journal} {\bibinfo  {journal} {Nano Letters}\ }\textbf {\bibinfo {volume} {21}},\ \bibinfo {pages} {3511} (\bibinfo {year} {2021})}\BibitemShut {NoStop}%
\bibitem [{\citenamefont {Wang}\ \emph {et~al.}(2019)\citenamefont {Wang}, \citenamefont {Gibertini}, \citenamefont {Dumcenco}, \citenamefont {Taniguchi}, \citenamefont {Watanabe}, \citenamefont {Giannini},\ and\ \citenamefont {Morpurgo}}]{3wang2019determining}%
  \BibitemOpen
  \bibfield  {author} {\bibinfo {author} {\bibfnamefont {Z.}~\bibnamefont {Wang}}, \bibinfo {author} {\bibfnamefont {M.}~\bibnamefont {Gibertini}}, \bibinfo {author} {\bibfnamefont {D.}~\bibnamefont {Dumcenco}}, \bibinfo {author} {\bibfnamefont {T.}~\bibnamefont {Taniguchi}}, \bibinfo {author} {\bibfnamefont {K.}~\bibnamefont {Watanabe}}, \bibinfo {author} {\bibfnamefont {E.}~\bibnamefont {Giannini}},\ and\ \bibinfo {author} {\bibfnamefont {A.~F.}\ \bibnamefont {Morpurgo}},\ }\bibfield  {title} {\bibinfo {title} {{Determining the phase diagram of atomically thin layered antiferromagnet CrCl$_{3}$}},\ }\href@noop {} {\bibfield  {journal} {\bibinfo  {journal} {Nature nanotechnology}\ }\textbf {\bibinfo {volume} {14}},\ \bibinfo {pages} {1116} (\bibinfo {year} {2019})}\BibitemShut {NoStop}%
\bibitem [{\citenamefont {DiScala}\ \emph {et~al.}(2024)\citenamefont {DiScala}, \citenamefont {Staros}, \citenamefont {de~la Torre}, \citenamefont {Lopez}, \citenamefont {Wong}, \citenamefont {Schulz}, \citenamefont {Barkowiak}, \citenamefont {Bisogni}, \citenamefont {Pelliciari}, \citenamefont {Rubenstein},\ and\ \citenamefont {Plumb}}]{3discala2024elucidating}%
  \BibitemOpen
  \bibfield  {author} {\bibinfo {author} {\bibfnamefont {M.~F.}\ \bibnamefont {DiScala}}, \bibinfo {author} {\bibfnamefont {D.}~\bibnamefont {Staros}}, \bibinfo {author} {\bibfnamefont {A.}~\bibnamefont {de~la Torre}}, \bibinfo {author} {\bibfnamefont {A.}~\bibnamefont {Lopez}}, \bibinfo {author} {\bibfnamefont {D.}~\bibnamefont {Wong}}, \bibinfo {author} {\bibfnamefont {C.}~\bibnamefont {Schulz}}, \bibinfo {author} {\bibfnamefont {M.}~\bibnamefont {Barkowiak}}, \bibinfo {author} {\bibfnamefont {V.}~\bibnamefont {Bisogni}}, \bibinfo {author} {\bibfnamefont {J.}~\bibnamefont {Pelliciari}}, \bibinfo {author} {\bibfnamefont {B.}~\bibnamefont {Rubenstein}},\ and\ \bibinfo {author} {\bibfnamefont {K.~W.}\ \bibnamefont {Plumb}},\ }\bibfield  {title} {\bibinfo {title} {{Elucidating the role of dimensionality on the electronic structure of the van der Waals antiferromagnet NiPS$_{3}$}},\ }\href@noop {} {\bibfield  {journal} {\bibinfo  {journal} {Advanced Physics Research}\ }\textbf {\bibinfo {volume} {3}},\ \bibinfo
  {pages} {2300096} (\bibinfo {year} {2024})}\BibitemShut {NoStop}%
\bibitem [{\citenamefont {Des~Cloizeaux}\ and\ \citenamefont {Pearson}(1962)}]{des1962spin}%
  \BibitemOpen
  \bibfield  {author} {\bibinfo {author} {\bibfnamefont {J.}~\bibnamefont {Des~Cloizeaux}}\ and\ \bibinfo {author} {\bibfnamefont {J.~J.}\ \bibnamefont {Pearson}},\ }\bibfield  {title} {\bibinfo {title} {Spin-wave spectrum of the antiferromagnetic linear chain},\ }\href@noop {} {\bibfield  {journal} {\bibinfo  {journal} {Physical Review}\ }\textbf {\bibinfo {volume} {128}},\ \bibinfo {pages} {2131} (\bibinfo {year} {1962})}\BibitemShut {NoStop}%
\bibitem [{\citenamefont {Nagler}\ \emph {et~al.}(1991)\citenamefont {Nagler}, \citenamefont {Tennant}, \citenamefont {Cowley}, \citenamefont {Perring},\ and\ \citenamefont {Satija}}]{nagler1991spin}%
  \BibitemOpen
  \bibfield  {author} {\bibinfo {author} {\bibfnamefont {S.}~\bibnamefont {Nagler}}, \bibinfo {author} {\bibfnamefont {D.}~\bibnamefont {Tennant}}, \bibinfo {author} {\bibfnamefont {R.}~\bibnamefont {Cowley}}, \bibinfo {author} {\bibfnamefont {T.}~\bibnamefont {Perring}},\ and\ \bibinfo {author} {\bibfnamefont {S.}~\bibnamefont {Satija}},\ }\bibfield  {title} {\bibinfo {title} {{Spin dynamics in the quantum antiferromagnetic chain compound KCuF$_{3}$}},\ }\href@noop {} {\bibfield  {journal} {\bibinfo  {journal} {Physical Review B}\ }\textbf {\bibinfo {volume} {44}},\ \bibinfo {pages} {12361} (\bibinfo {year} {1991})}\BibitemShut {NoStop}%
\bibitem [{\citenamefont {Reigrotzki}\ \emph {et~al.}(1994)\citenamefont {Reigrotzki}, \citenamefont {Tsunetsugu},\ and\ \citenamefont {Rice}}]{reigrotzki1994strong}%
  \BibitemOpen
  \bibfield  {author} {\bibinfo {author} {\bibfnamefont {M.}~\bibnamefont {Reigrotzki}}, \bibinfo {author} {\bibfnamefont {H.}~\bibnamefont {Tsunetsugu}},\ and\ \bibinfo {author} {\bibfnamefont {T.}~\bibnamefont {Rice}},\ }\bibfield  {title} {\bibinfo {title} {{Strong-coupling expansions for antiferromagnetic Heisenberg spin-one-half ladders}},\ }\href@noop {} {\bibfield  {journal} {\bibinfo  {journal} {Journal of Physics: Condensed Matter}\ }\textbf {\bibinfo {volume} {6}},\ \bibinfo {pages} {9235} (\bibinfo {year} {1994})}\BibitemShut {NoStop}%
\bibitem [{\citenamefont {Chen}\ \emph {et~al.}(1992)\citenamefont {Chen}, \citenamefont {Ding},\ and\ \citenamefont {Goddard~III}}]{chen1992elementary}%
  \BibitemOpen
  \bibfield  {author} {\bibinfo {author} {\bibfnamefont {G.}~\bibnamefont {Chen}}, \bibinfo {author} {\bibfnamefont {H.-Q.}\ \bibnamefont {Ding}},\ and\ \bibinfo {author} {\bibfnamefont {W.~A.}\ \bibnamefont {Goddard~III}},\ }\bibfield  {title} {\bibinfo {title} {{Elementary excitations for the two-dimensional quantum Heisenberg antiferromagnet}},\ }\href@noop {} {\bibfield  {journal} {\bibinfo  {journal} {Physical Review B}\ }\textbf {\bibinfo {volume} {46}},\ \bibinfo {pages} {2933} (\bibinfo {year} {1992})}\BibitemShut {NoStop}%
\bibitem [{\citenamefont {Richter}\ and\ \citenamefont {Schulenburg}(2010)}]{richter2010spin}%
  \BibitemOpen
  \bibfield  {author} {\bibinfo {author} {\bibfnamefont {J.}~\bibnamefont {Richter}}\ and\ \bibinfo {author} {\bibfnamefont {J.}~\bibnamefont {Schulenburg}},\ }\bibfield  {title} {\bibinfo {title} {{The spin-1/2 J1--J2 Heisenberg antiferromagnet on the square lattice: Exact diagonalization for N= 40 spins}},\ }\href@noop {} {\bibfield  {journal} {\bibinfo  {journal} {The European Physical Journal B}\ }\textbf {\bibinfo {volume} {73}},\ \bibinfo {pages} {117} (\bibinfo {year} {2010})}\BibitemShut {NoStop}%
\bibitem [{\citenamefont {Wang}\ and\ \citenamefont {Sandvik}(2010)}]{wang2010low}%
  \BibitemOpen
  \bibfield  {author} {\bibinfo {author} {\bibfnamefont {L.}~\bibnamefont {Wang}}\ and\ \bibinfo {author} {\bibfnamefont {A.~W.}\ \bibnamefont {Sandvik}},\ }\bibfield  {title} {\bibinfo {title} {{Low-energy excitations of two-dimensional diluted Heisenberg quantum antiferromagnets}},\ }\href@noop {} {\bibfield  {journal} {\bibinfo  {journal} {{Physical Review B--Condensed Matter and Materials Physics}}\ }\textbf {\bibinfo {volume} {81}},\ \bibinfo {pages} {054417} (\bibinfo {year} {2010})}\BibitemShut {NoStop}%
\bibitem [{\citenamefont {Weihong}(1997)}]{weihong1997various}%
  \BibitemOpen
  \bibfield  {author} {\bibinfo {author} {\bibfnamefont {Z.}~\bibnamefont {Weihong}},\ }\bibfield  {title} {\bibinfo {title} {{Various series expansions for the bilayer S= 1/ 2 Heisenberg antiferromagnet}},\ }\href@noop {} {\bibfield  {journal} {\bibinfo  {journal} {Physical Review B}\ }\textbf {\bibinfo {volume} {55}},\ \bibinfo {pages} {12267} (\bibinfo {year} {1997})}\BibitemShut {NoStop}%
\bibitem [{\citenamefont {Wang}\ \emph {et~al.}(2006)\citenamefont {Wang}, \citenamefont {Beach},\ and\ \citenamefont {Sandvik}}]{wang2006high}%
  \BibitemOpen
  \bibfield  {author} {\bibinfo {author} {\bibfnamefont {L.}~\bibnamefont {Wang}}, \bibinfo {author} {\bibfnamefont {K.}~\bibnamefont {Beach}},\ and\ \bibinfo {author} {\bibfnamefont {A.~W.}\ \bibnamefont {Sandvik}},\ }\bibfield  {title} {\bibinfo {title} {{High-precision finite-size scaling analysis of the quantum-critical point of S= 1/ 2 Heisenberg antiferromagnetic bilayers}},\ }\href@noop {} {\bibfield  {journal} {\bibinfo  {journal} {Physical Review B--Condensed Matter and Materials Physics}\ }\textbf {\bibinfo {volume} {73}},\ \bibinfo {pages} {014431} (\bibinfo {year} {2006})}\BibitemShut {NoStop}%
\bibitem [{\citenamefont {Ma}\ \emph {et~al.}(2015)\citenamefont {Ma}, \citenamefont {Sandvik},\ and\ \citenamefont {Yao}}]{ma2015mott}%
  \BibitemOpen
  \bibfield  {author} {\bibinfo {author} {\bibfnamefont {N.-S.}\ \bibnamefont {Ma}}, \bibinfo {author} {\bibfnamefont {A.~W.}\ \bibnamefont {Sandvik}},\ and\ \bibinfo {author} {\bibfnamefont {D.-X.}\ \bibnamefont {Yao}},\ }\bibfield  {title} {\bibinfo {title} {{Mott glass phase in a diluted bilayer Heisenberg quantum antiferromagnet}},\ }\href@noop {} {\bibfield  {journal} {\bibinfo  {journal} {Journal of Physics: Conference Series}\ }\textbf {\bibinfo {volume} {640}},\ \bibinfo {pages} {012045} (\bibinfo {year} {2015})}\BibitemShut {NoStop}%
\bibitem [{\citenamefont {Sasago}\ \emph {et~al.}(1997)\citenamefont {Sasago}, \citenamefont {Uchinokura}, \citenamefont {Zheludev},\ and\ \citenamefont {Shirane}}]{sasago1997temperature}%
  \BibitemOpen
  \bibfield  {author} {\bibinfo {author} {\bibfnamefont {Y.}~\bibnamefont {Sasago}}, \bibinfo {author} {\bibfnamefont {K.}~\bibnamefont {Uchinokura}}, \bibinfo {author} {\bibfnamefont {A.}~\bibnamefont {Zheludev}},\ and\ \bibinfo {author} {\bibfnamefont {G.}~\bibnamefont {Shirane}},\ }\bibfield  {title} {\bibinfo {title} {{Temperature-dependent spin gap and singlet ground state in BaCuSi$_{2}$O$_{6}$}},\ }\href@noop {} {\bibfield  {journal} {\bibinfo  {journal} {Physcial Review B}\ }\textbf {\bibinfo {volume} {55}},\ \bibinfo {pages} {8357} (\bibinfo {year} {1997})}\BibitemShut {NoStop}%
\bibitem [{\citenamefont {Leuenberger}\ \emph {et~al.}(1984)\citenamefont {Leuenberger}, \citenamefont {Stebler}, \citenamefont {G{\"u}del}, \citenamefont {Furrer}, \citenamefont {Feile},\ and\ \citenamefont {Kjems}}]{leuenberger1984spin}%
  \BibitemOpen
  \bibfield  {author} {\bibinfo {author} {\bibfnamefont {B.}~\bibnamefont {Leuenberger}}, \bibinfo {author} {\bibfnamefont {A.}~\bibnamefont {Stebler}}, \bibinfo {author} {\bibfnamefont {H.}~\bibnamefont {G{\"u}del}}, \bibinfo {author} {\bibfnamefont {A.}~\bibnamefont {Furrer}}, \bibinfo {author} {\bibfnamefont {R.}~\bibnamefont {Feile}},\ and\ \bibinfo {author} {\bibfnamefont {J.}~\bibnamefont {Kjems}},\ }\bibfield  {title} {\bibinfo {title} {{Spin dynamics of an isotropic singlet-ground-state antiferromagnet with alternating strong and weak interactions: An inelastic-neutron-scattering study of the dimer compound Cs$_{3}$Cr$_{2}$Br$_{9}$}},\ }\href@noop {} {\bibfield  {journal} {\bibinfo  {journal} {Physical Review B}\ }\textbf {\bibinfo {volume} {30}},\ \bibinfo {pages} {6300} (\bibinfo {year} {1984})}\BibitemShut {NoStop}%
\bibitem [{\citenamefont {Sun}\ \emph {et~al.}(2023)\citenamefont {Sun}, \citenamefont {Huo}, \citenamefont {Hu}, \citenamefont {Li}, \citenamefont {Liu}, \citenamefont {Han}, \citenamefont {Tang}, \citenamefont {Mao}, \citenamefont {Yang}, \citenamefont {Wang}, \citenamefont {Cheng}, \citenamefont {Yao}, \citenamefont {Zhang},\ and\ \citenamefont {Wang}}]{sun2023signatures}%
  \BibitemOpen
  \bibfield  {author} {\bibinfo {author} {\bibfnamefont {H.}~\bibnamefont {Sun}}, \bibinfo {author} {\bibfnamefont {M.}~\bibnamefont {Huo}}, \bibinfo {author} {\bibfnamefont {X.}~\bibnamefont {Hu}}, \bibinfo {author} {\bibfnamefont {J.}~\bibnamefont {Li}}, \bibinfo {author} {\bibfnamefont {Z.}~\bibnamefont {Liu}}, \bibinfo {author} {\bibfnamefont {Y.}~\bibnamefont {Han}}, \bibinfo {author} {\bibfnamefont {L.}~\bibnamefont {Tang}}, \bibinfo {author} {\bibfnamefont {Z.}~\bibnamefont {Mao}}, \bibinfo {author} {\bibfnamefont {P.}~\bibnamefont {Yang}}, \bibinfo {author} {\bibfnamefont {B.}~\bibnamefont {Wang}}, \bibinfo {author} {\bibfnamefont {J.}~\bibnamefont {Cheng}}, \bibinfo {author} {\bibfnamefont {D.-X.}\ \bibnamefont {Yao}}, \bibinfo {author} {\bibfnamefont {G.-M.}\ \bibnamefont {Zhang}},\ and\ \bibinfo {author} {\bibfnamefont {M.}~\bibnamefont {Wang}},\ }\bibfield  {title} {\bibinfo {title} {Signatures of superconductivity near 80 k in a nickelate under high pressure},\ }\href@noop {} {\bibfield
  {journal} {\bibinfo  {journal} {Nature}\ }\textbf {\bibinfo {volume} {621}},\ \bibinfo {pages} {493} (\bibinfo {year} {2023})}\BibitemShut {NoStop}%
\bibitem [{\citenamefont {Luo}\ \emph {et~al.}(2023)\citenamefont {Luo}, \citenamefont {Hu}, \citenamefont {Wang}, \citenamefont {W{\'u}},\ and\ \citenamefont {Yao}}]{luo2023bilayer}%
  \BibitemOpen
  \bibfield  {author} {\bibinfo {author} {\bibfnamefont {Z.}~\bibnamefont {Luo}}, \bibinfo {author} {\bibfnamefont {X.}~\bibnamefont {Hu}}, \bibinfo {author} {\bibfnamefont {M.}~\bibnamefont {Wang}}, \bibinfo {author} {\bibfnamefont {W.}~\bibnamefont {W{\'u}}},\ and\ \bibinfo {author} {\bibfnamefont {D.-X.}\ \bibnamefont {Yao}},\ }\bibfield  {title} {\bibinfo {title} {{Bilayer two-orbital model of L a$_{3}$N i$_{2}$O$_{7}$ under pressure}},\ }\href@noop {} {\bibfield  {journal} {\bibinfo  {journal} {Physical review letters}\ }\textbf {\bibinfo {volume} {131}},\ \bibinfo {pages} {126001} (\bibinfo {year} {2023})}\BibitemShut {NoStop}%
\bibitem [{\citenamefont {Li}\ \emph {et~al.}(2019)\citenamefont {Li}, \citenamefont {Li}, \citenamefont {Du}, \citenamefont {Wang}, \citenamefont {Gu}, \citenamefont {Zhang}, \citenamefont {He}, \citenamefont {Duan},\ and\ \citenamefont {Xu}}]{li2019intrinsic}%
  \BibitemOpen
  \bibfield  {author} {\bibinfo {author} {\bibfnamefont {J.}~\bibnamefont {Li}}, \bibinfo {author} {\bibfnamefont {Y.}~\bibnamefont {Li}}, \bibinfo {author} {\bibfnamefont {S.}~\bibnamefont {Du}}, \bibinfo {author} {\bibfnamefont {Z.}~\bibnamefont {Wang}}, \bibinfo {author} {\bibfnamefont {B.-L.}\ \bibnamefont {Gu}}, \bibinfo {author} {\bibfnamefont {S.-C.}\ \bibnamefont {Zhang}}, \bibinfo {author} {\bibfnamefont {K.}~\bibnamefont {He}}, \bibinfo {author} {\bibfnamefont {W.}~\bibnamefont {Duan}},\ and\ \bibinfo {author} {\bibfnamefont {Y.}~\bibnamefont {Xu}},\ }\bibfield  {title} {\bibinfo {title} {{Intrinsic magnetic topological insulators in van der Waals layered MnBi$_{2}$Te$_{4}$-family materials}},\ }\href@noop {} {\bibfield  {journal} {\bibinfo  {journal} {Science Advances}\ }\textbf {\bibinfo {volume} {5}},\ \bibinfo {pages} {eaaw5685} (\bibinfo {year} {2019})}\BibitemShut {NoStop}%
\bibitem [{\citenamefont {Cheng}\ \emph {et~al.}(2022)\citenamefont {Cheng}, \citenamefont {Li}, \citenamefont {Xiong}, \citenamefont {Wu}, \citenamefont {Sandvik},\ and\ \citenamefont {Yao}}]{cheng2022fractional}%
  \BibitemOpen
  \bibfield  {author} {\bibinfo {author} {\bibfnamefont {J.-Q.}\ \bibnamefont {Cheng}}, \bibinfo {author} {\bibfnamefont {J.}~\bibnamefont {Li}}, \bibinfo {author} {\bibfnamefont {Z.}~\bibnamefont {Xiong}}, \bibinfo {author} {\bibfnamefont {H.-Q.}\ \bibnamefont {Wu}}, \bibinfo {author} {\bibfnamefont {A.~W.}\ \bibnamefont {Sandvik}},\ and\ \bibinfo {author} {\bibfnamefont {D.-X.}\ \bibnamefont {Yao}},\ }\bibfield  {title} {\bibinfo {title} {Fractional and composite excitations of antiferromagnetic quantum spin trimer chains},\ }\href@noop {} {\bibfield  {journal} {\bibinfo  {journal} {npj Quantum Materials}\ }\textbf {\bibinfo {volume} {7}},\ \bibinfo {pages} {3} (\bibinfo {year} {2022})}\BibitemShut {NoStop}%
\bibitem [{\citenamefont {Chang}\ \emph {et~al.}(2024)\citenamefont {Chang}, \citenamefont {Cheng}, \citenamefont {Shao}, \citenamefont {Yao},\ and\ \citenamefont {Wu}}]{chang2024magnon}%
  \BibitemOpen
  \bibfield  {author} {\bibinfo {author} {\bibfnamefont {Y.-Y.}\ \bibnamefont {Chang}}, \bibinfo {author} {\bibfnamefont {J.-Q.}\ \bibnamefont {Cheng}}, \bibinfo {author} {\bibfnamefont {H.}~\bibnamefont {Shao}}, \bibinfo {author} {\bibfnamefont {D.-X.}\ \bibnamefont {Yao}},\ and\ \bibinfo {author} {\bibfnamefont {H.-Q.}\ \bibnamefont {Wu}},\ }\bibfield  {title} {\bibinfo {title} {{Magnon, doublon and quarton excitations in 2D S= 1/2 trimerized Heisenberg models}},\ }\href@noop {} {\bibfield  {journal} {\bibinfo  {journal} {Frontiers of Physics}\ }\textbf {\bibinfo {volume} {19}},\ \bibinfo {pages} {63202} (\bibinfo {year} {2024})}\BibitemShut {NoStop}%
\bibitem [{\citenamefont {Hu}\ \emph {et~al.}(2013)\citenamefont {Hu}, \citenamefont {Shi}, \citenamefont {Jia}, \citenamefont {Liu}, \citenamefont {Wu},\ and\ \citenamefont {Du}}]{hu2013exchange}%
  \BibitemOpen
  \bibfield  {author} {\bibinfo {author} {\bibfnamefont {Y.}~\bibnamefont {Hu}}, \bibinfo {author} {\bibfnamefont {F.}~\bibnamefont {Shi}}, \bibinfo {author} {\bibfnamefont {N.}~\bibnamefont {Jia}}, \bibinfo {author} {\bibfnamefont {Y.}~\bibnamefont {Liu}}, \bibinfo {author} {\bibfnamefont {H.}~\bibnamefont {Wu}},\ and\ \bibinfo {author} {\bibfnamefont {A.}~\bibnamefont {Du}},\ }\bibfield  {title} {\bibinfo {title} {Exchange bias and its propagation in ferromagnetic/antiferromagnetic/ferromagnetic trilayers},\ }\href@noop {} {\bibfield  {journal} {\bibinfo  {journal} {Journal of Applied Physics}\ }\textbf {\bibinfo {volume} {114}} (\bibinfo {year} {2013})}\BibitemShut {NoStop}%
\bibitem [{\citenamefont {Chatterjee}\ \emph {et~al.}(2022)\citenamefont {Chatterjee}, \citenamefont {Wang}, \citenamefont {Berg},\ and\ \citenamefont {Zaletel}}]{chatterjee2022inter}%
  \BibitemOpen
  \bibfield  {author} {\bibinfo {author} {\bibfnamefont {S.}~\bibnamefont {Chatterjee}}, \bibinfo {author} {\bibfnamefont {T.}~\bibnamefont {Wang}}, \bibinfo {author} {\bibfnamefont {E.}~\bibnamefont {Berg}},\ and\ \bibinfo {author} {\bibfnamefont {M.~P.}\ \bibnamefont {Zaletel}},\ }\bibfield  {title} {\bibinfo {title} {Inter-valley coherent order and isospin fluctuation mediated superconductivity in rhombohedral trilayer graphene},\ }\href@noop {} {\bibfield  {journal} {\bibinfo  {journal} {Nature communications}\ }\textbf {\bibinfo {volume} {13}},\ \bibinfo {pages} {6013} (\bibinfo {year} {2022})}\BibitemShut {NoStop}%
\bibitem [{\citenamefont {Weber}\ \emph {et~al.}(2022)\citenamefont {Weber}, \citenamefont {Honecker}, \citenamefont {Normand}, \citenamefont {Corboz}, \citenamefont {Mila},\ and\ \citenamefont {Wessel}}]{weber2022quantum}%
  \BibitemOpen
  \bibfield  {author} {\bibinfo {author} {\bibfnamefont {L.}~\bibnamefont {Weber}}, \bibinfo {author} {\bibfnamefont {A.}~\bibnamefont {Honecker}}, \bibinfo {author} {\bibfnamefont {B.}~\bibnamefont {Normand}}, \bibinfo {author} {\bibfnamefont {P.}~\bibnamefont {Corboz}}, \bibinfo {author} {\bibfnamefont {F.}~\bibnamefont {Mila}},\ and\ \bibinfo {author} {\bibfnamefont {S.}~\bibnamefont {Wessel}},\ }\bibfield  {title} {\bibinfo {title} {{Quantum Monte Carlo simulations in the trimer basis: first-order transitions and thermal critical points in frustrated trilayer magnets}},\ }\href@noop {} {\bibfield  {journal} {\bibinfo  {journal} {SciPost Physics}\ }\textbf {\bibinfo {volume} {12}},\ \bibinfo {pages} {054} (\bibinfo {year} {2022})}\BibitemShut {NoStop}%
\bibitem [{\citenamefont {Roth}\ \emph {et~al.}(1987)\citenamefont {Roth}, \citenamefont {Renker}, \citenamefont {Heger}, \citenamefont {Hervieu}, \citenamefont {Domenges},\ and\ \citenamefont {Raveau}}]{roth1987structure}%
  \BibitemOpen
  \bibfield  {author} {\bibinfo {author} {\bibfnamefont {G.}~\bibnamefont {Roth}}, \bibinfo {author} {\bibfnamefont {B.}~\bibnamefont {Renker}}, \bibinfo {author} {\bibfnamefont {G.}~\bibnamefont {Heger}}, \bibinfo {author} {\bibfnamefont {M.}~\bibnamefont {Hervieu}}, \bibinfo {author} {\bibfnamefont {B.}~\bibnamefont {Domenges}},\ and\ \bibinfo {author} {\bibfnamefont {B.}~\bibnamefont {Raveau}},\ }\bibfield  {title} {\bibinfo {title} {{On the structure of non-superconducting YBa$_{2}$Cu$_{3}$O$_{6+\epsilon}$}},\ }\href@noop {} {\bibfield  {journal} {\bibinfo  {journal} {Zeitschrift f{\"u}r Physik B Condensed Matter}\ }\textbf {\bibinfo {volume} {69}},\ \bibinfo {pages} {53} (\bibinfo {year} {1987})}\BibitemShut {NoStop}%
\bibitem [{\citenamefont {Shamray}\ \emph {et~al.}(2009)\citenamefont {Shamray}, \citenamefont {Mikhailova},\ and\ \citenamefont {Mitin}}]{shamray2009crystal}%
  \BibitemOpen
  \bibfield  {author} {\bibinfo {author} {\bibfnamefont {V.}~\bibnamefont {Shamray}}, \bibinfo {author} {\bibfnamefont {A.}~\bibnamefont {Mikhailova}},\ and\ \bibinfo {author} {\bibfnamefont {A.}~\bibnamefont {Mitin}},\ }\bibfield  {title} {\bibinfo {title} {{Crystal structure and superconductivity of Bi-2223}},\ }\href@noop {} {\bibfield  {journal} {\bibinfo  {journal} {Crystallography Reports}\ }\textbf {\bibinfo {volume} {54}},\ \bibinfo {pages} {584} (\bibinfo {year} {2009})}\BibitemShut {NoStop}%
\bibitem [{\citenamefont {Wang}\ and\ \citenamefont {Sanyal}(2021)}]{wang2021systematic}%
  \BibitemOpen
  \bibfield  {author} {\bibinfo {author} {\bibfnamefont {D.}~\bibnamefont {Wang}}\ and\ \bibinfo {author} {\bibfnamefont {B.}~\bibnamefont {Sanyal}},\ }\bibfield  {title} {\bibinfo {title} {{Systematic study of monolayer to trilayer CrI$_{3}$: Stacking sequence dependence of electronic structure and magnetism}},\ }\href@noop {} {\bibfield  {journal} {\bibinfo  {journal} {The Journal of Physical Chemistry C}\ }\textbf {\bibinfo {volume} {125}},\ \bibinfo {pages} {18467} (\bibinfo {year} {2021})}\BibitemShut {NoStop}%
\bibitem [{\citenamefont {Zhang}\ \emph {et~al.}(2020)\citenamefont {Zhang}, \citenamefont {Phelan}, \citenamefont {Botana}, \citenamefont {Chen}, \citenamefont {Zheng}, \citenamefont {Krogstad}, \citenamefont {Wang}, \citenamefont {Qiu}, \citenamefont {Rodriguez-Rivera}, \citenamefont {Osborn}, \citenamefont {Rosenkranz}, \citenamefont {Norman},\ and\ \citenamefont {Mitchell}}]{zhang2020intertwined}%
  \BibitemOpen
  \bibfield  {author} {\bibinfo {author} {\bibfnamefont {J.}~\bibnamefont {Zhang}}, \bibinfo {author} {\bibfnamefont {D.}~\bibnamefont {Phelan}}, \bibinfo {author} {\bibfnamefont {A.}~\bibnamefont {Botana}}, \bibinfo {author} {\bibfnamefont {Y.-S.}\ \bibnamefont {Chen}}, \bibinfo {author} {\bibfnamefont {H.}~\bibnamefont {Zheng}}, \bibinfo {author} {\bibfnamefont {M.}~\bibnamefont {Krogstad}}, \bibinfo {author} {\bibfnamefont {S.~G.}\ \bibnamefont {Wang}}, \bibinfo {author} {\bibfnamefont {Y.}~\bibnamefont {Qiu}}, \bibinfo {author} {\bibfnamefont {J.}~\bibnamefont {Rodriguez-Rivera}}, \bibinfo {author} {\bibfnamefont {R.}~\bibnamefont {Osborn}}, \bibinfo {author} {\bibfnamefont {S.}~\bibnamefont {Rosenkranz}}, \bibinfo {author} {\bibfnamefont {M.~R.}\ \bibnamefont {Norman}},\ and\ \bibinfo {author} {\bibfnamefont {J.~F.}\ \bibnamefont {Mitchell}},\ }\bibfield  {title} {\bibinfo {title} {Intertwined density waves in a metallic nickelate},\ }\href@noop {} {\bibfield  {journal} {\bibinfo  {journal} {Nature
  communications}\ }\textbf {\bibinfo {volume} {11}},\ \bibinfo {pages} {6003} (\bibinfo {year} {2020})}\BibitemShut {NoStop}%
\bibitem [{\citenamefont {Qin}\ \emph {et~al.}(2024)\citenamefont {Qin}, \citenamefont {Wang},\ and\ \citenamefont {Yang}}]{qin2024frustrated}%
  \BibitemOpen
  \bibfield  {author} {\bibinfo {author} {\bibfnamefont {Q.}~\bibnamefont {Qin}}, \bibinfo {author} {\bibfnamefont {J.}~\bibnamefont {Wang}},\ and\ \bibinfo {author} {\bibfnamefont {Y.}~\bibnamefont {Yang}},\ }\bibfield  {title} {\bibinfo {title} {{Frustrated Superconductivity and Intrinsic Reduction of Tc in Trilayer Nickelate}},\ }\href@noop {} {\bibfield  {journal} {\bibinfo  {journal} {The Innovation Materials}\ } (\bibinfo {year} {2024})}\BibitemShut {NoStop}%
\bibitem [{\citenamefont {Yao}\ and\ \citenamefont {Carlson}(2008)}]{yao2008magnetic}%
  \BibitemOpen
  \bibfield  {author} {\bibinfo {author} {\bibfnamefont {D.-X.}\ \bibnamefont {Yao}}\ and\ \bibinfo {author} {\bibfnamefont {E.~W.}\ \bibnamefont {Carlson}},\ }\bibfield  {title} {\bibinfo {title} {{Magnetic excitations in the high-T$_{c}$ iron pnictides}},\ }\href@noop {} {\bibfield  {journal} {\bibinfo  {journal} {Physical Review B--Condensed Matter and Materials Physics}\ }\textbf {\bibinfo {volume} {78}},\ \bibinfo {pages} {052507} (\bibinfo {year} {2008})}\BibitemShut {NoStop}%
\bibitem [{\citenamefont {Tan}\ and\ \citenamefont {Yao}(2023)}]{tan2023spin}%
  \BibitemOpen
  \bibfield  {author} {\bibinfo {author} {\bibfnamefont {Y.}~\bibnamefont {Tan}}\ and\ \bibinfo {author} {\bibfnamefont {D.-X.}\ \bibnamefont {Yao}},\ }\bibfield  {title} {\bibinfo {title} {Spin waves and phase transition on a magnetically frustrated square lattice with long-range interactions},\ }\href@noop {} {\bibfield  {journal} {\bibinfo  {journal} {Frontiers of Physics}\ }\textbf {\bibinfo {volume} {18}},\ \bibinfo {pages} {33309} (\bibinfo {year} {2023})}\BibitemShut {NoStop}%
\bibitem [{\citenamefont {Carlson}\ \emph {et~al.}(2004)\citenamefont {Carlson}, \citenamefont {Yao},\ and\ \citenamefont {Campbell}}]{carlson2004spin}%
  \BibitemOpen
  \bibfield  {author} {\bibinfo {author} {\bibfnamefont {E.}~\bibnamefont {Carlson}}, \bibinfo {author} {\bibfnamefont {D.}~\bibnamefont {Yao}},\ and\ \bibinfo {author} {\bibfnamefont {D.}~\bibnamefont {Campbell}},\ }\bibfield  {title} {\bibinfo {title} {Spin waves in striped phases},\ }\href@noop {} {\bibfield  {journal} {\bibinfo  {journal} {Physical Review B--Condensed Matter and Materials Physics}\ }\textbf {\bibinfo {volume} {70}},\ \bibinfo {pages} {064505} (\bibinfo {year} {2004})}\BibitemShut {NoStop}%
\bibitem [{\citenamefont {Shao}\ \emph {et~al.}(2017)\citenamefont {Shao}, \citenamefont {Qin}, \citenamefont {Capponi}, \citenamefont {Chesi}, \citenamefont {Meng},\ and\ \citenamefont {Sandvik}}]{shao2017nearly}%
  \BibitemOpen
  \bibfield  {author} {\bibinfo {author} {\bibfnamefont {H.}~\bibnamefont {Shao}}, \bibinfo {author} {\bibfnamefont {Y.~Q.}\ \bibnamefont {Qin}}, \bibinfo {author} {\bibfnamefont {S.}~\bibnamefont {Capponi}}, \bibinfo {author} {\bibfnamefont {S.}~\bibnamefont {Chesi}}, \bibinfo {author} {\bibfnamefont {Z.~Y.}\ \bibnamefont {Meng}},\ and\ \bibinfo {author} {\bibfnamefont {A.~W.}\ \bibnamefont {Sandvik}},\ }\bibfield  {title} {\bibinfo {title} {{Nearly deconfined spinon excitations in the square-lattice spin-1/2 Heisenberg antiferromagnet}},\ }\href@noop {} {\bibfield  {journal} {\bibinfo  {journal} {Physical Review X}\ }\textbf {\bibinfo {volume} {7}},\ \bibinfo {pages} {041072} (\bibinfo {year} {2017})}\BibitemShut {NoStop}%
\bibitem [{\citenamefont {Dalla~Piazza}\ \emph {et~al.}(2015)\citenamefont {Dalla~Piazza}, \citenamefont {Mourigal}, \citenamefont {Christensen}, \citenamefont {Nilsen}, \citenamefont {Tregenna-Piggott}, \citenamefont {Perring}, \citenamefont {Enderle}, \citenamefont {McMorrow}, \citenamefont {Ivanov},\ and\ \citenamefont {R{\o}nnow}}]{dalla2015fractional}%
  \BibitemOpen
  \bibfield  {author} {\bibinfo {author} {\bibfnamefont {B.}~\bibnamefont {Dalla~Piazza}}, \bibinfo {author} {\bibfnamefont {M.}~\bibnamefont {Mourigal}}, \bibinfo {author} {\bibfnamefont {N.~B.}\ \bibnamefont {Christensen}}, \bibinfo {author} {\bibfnamefont {G.}~\bibnamefont {Nilsen}}, \bibinfo {author} {\bibfnamefont {P.}~\bibnamefont {Tregenna-Piggott}}, \bibinfo {author} {\bibfnamefont {T.}~\bibnamefont {Perring}}, \bibinfo {author} {\bibfnamefont {M.}~\bibnamefont {Enderle}}, \bibinfo {author} {\bibfnamefont {D.~F.}\ \bibnamefont {McMorrow}}, \bibinfo {author} {\bibfnamefont {D.}~\bibnamefont {Ivanov}},\ and\ \bibinfo {author} {\bibfnamefont {H.~M.}\ \bibnamefont {R{\o}nnow}},\ }\bibfield  {title} {\bibinfo {title} {Fractional excitations in the square-lattice quantum antiferromagnet},\ }\href@noop {} {\bibfield  {journal} {\bibinfo  {journal} {Nature physics}\ }\textbf {\bibinfo {volume} {11}},\ \bibinfo {pages} {62} (\bibinfo {year} {2015})}\BibitemShut {NoStop}%
\bibitem [{\citenamefont {Headings}\ \emph {et~al.}(2010)\citenamefont {Headings}, \citenamefont {Hayden}, \citenamefont {Coldea},\ and\ \citenamefont {Perring}}]{headings2010anomalous}%
  \BibitemOpen
  \bibfield  {author} {\bibinfo {author} {\bibfnamefont {N.}~\bibnamefont {Headings}}, \bibinfo {author} {\bibfnamefont {S.}~\bibnamefont {Hayden}}, \bibinfo {author} {\bibfnamefont {R.}~\bibnamefont {Coldea}},\ and\ \bibinfo {author} {\bibfnamefont {T.}~\bibnamefont {Perring}},\ }\bibfield  {title} {\bibinfo {title} {{Anomalous high-energy spin excitations in the high-T c superconductor-parent antiferromagnet La$_{2}$CuO$_{4}$}},\ }\href@noop {} {\bibfield  {journal} {\bibinfo  {journal} {Physical review letters}\ }\textbf {\bibinfo {volume} {105}},\ \bibinfo {pages} {247001} (\bibinfo {year} {2010})}\BibitemShut {NoStop}%
\bibitem [{\citenamefont {Abbas}\ \emph {et~al.}(2012)\citenamefont {Abbas}, \citenamefont {Abbas},\ and\ \citenamefont {Mahdi}}]{abbas2012effect}%
  \BibitemOpen
  \bibfield  {author} {\bibinfo {author} {\bibfnamefont {L.}~\bibnamefont {Abbas}}, \bibinfo {author} {\bibfnamefont {M.}~\bibnamefont {Abbas}},\ and\ \bibinfo {author} {\bibfnamefont {O.}~\bibnamefont {Mahdi}},\ }\bibfield  {title} {\bibinfo {title} {{Effect of Sintering Temperature on Bi$_{2-x}$Cu$_{x}$Pb$_{0.3}$Sr$_{2}$Ca$_{2}$Cu$_{3}$O$_{10+\delta}$ Superconductors}},\ }\href@noop {} {\bibfield  {journal} {\bibinfo  {journal} {Al-Nahrain Journal of Science}\ }\textbf {\bibinfo {volume} {15}},\ \bibinfo {pages} {113} (\bibinfo {year} {2012})}\BibitemShut {NoStop}%
\bibitem [{\citenamefont {Ali}\ and\ \citenamefont {Jasim}(2024)}]{ali2024study}%
  \BibitemOpen
  \bibfield  {author} {\bibinfo {author} {\bibfnamefont {F.~W.}\ \bibnamefont {Ali}}\ and\ \bibinfo {author} {\bibfnamefont {K.~A.}\ \bibnamefont {Jasim}},\ }\bibfield  {title} {\bibinfo {title} {{Study of Some Physical Properties of the Superconducting Compound PbBa$_{2}$Ca$_{2}$Cu$_{3}$O$_{8+\delta}$}},\ }\href@noop {} {\bibfield  {journal} {\bibinfo  {journal} {Ibn AL-Haitham Journal For Pure and Applied Sciences}\ }\textbf {\bibinfo {volume} {37}},\ \bibinfo {pages} {168} (\bibinfo {year} {2024})}\BibitemShut {NoStop}%
\bibitem [{\citenamefont {Adams}\ \emph {et~al.}(2019)\citenamefont {Adams}, \citenamefont {Pritchard},\ and\ \citenamefont {Shaffer}}]{adams2019rydberg}%
  \BibitemOpen
  \bibfield  {author} {\bibinfo {author} {\bibfnamefont {C.~S.}\ \bibnamefont {Adams}}, \bibinfo {author} {\bibfnamefont {J.~D.}\ \bibnamefont {Pritchard}},\ and\ \bibinfo {author} {\bibfnamefont {J.~P.}\ \bibnamefont {Shaffer}},\ }\bibfield  {title} {\bibinfo {title} {Rydberg atom quantum technologies},\ }\href@noop {} {\bibfield  {journal} {\bibinfo  {journal} {Journal of Physics B: Atomic, Molecular and Optical Physics}\ }\textbf {\bibinfo {volume} {53}},\ \bibinfo {pages} {012002} (\bibinfo {year} {2019})}\BibitemShut {NoStop}%
\bibitem [{\citenamefont {Wirth}\ \emph {et~al.}(2011)\citenamefont {Wirth}, \citenamefont {{\"O}lschl{\"a}ger},\ and\ \citenamefont {Hemmerich}}]{wirth2011evidence}%
  \BibitemOpen
  \bibfield  {author} {\bibinfo {author} {\bibfnamefont {G.}~\bibnamefont {Wirth}}, \bibinfo {author} {\bibfnamefont {M.}~\bibnamefont {{\"O}lschl{\"a}ger}},\ and\ \bibinfo {author} {\bibfnamefont {A.}~\bibnamefont {Hemmerich}},\ }\bibfield  {title} {\bibinfo {title} {Evidence for orbital superfluidity in the p-band of a bipartite optical square lattice},\ }\href@noop {} {\bibfield  {journal} {\bibinfo  {journal} {Nature Physics}\ }\textbf {\bibinfo {volume} {7}},\ \bibinfo {pages} {147} (\bibinfo {year} {2011})}\BibitemShut {NoStop}%
\bibitem [{\citenamefont {Gross}\ and\ \citenamefont {Bloch}(2017)}]{gross2017quantum}%
  \BibitemOpen
  \bibfield  {author} {\bibinfo {author} {\bibfnamefont {C.}~\bibnamefont {Gross}}\ and\ \bibinfo {author} {\bibfnamefont {I.}~\bibnamefont {Bloch}},\ }\bibfield  {title} {\bibinfo {title} {Quantum simulations with ultracold atoms in optical lattices},\ }\href@noop {} {\bibfield  {journal} {\bibinfo  {journal} {Science}\ }\textbf {\bibinfo {volume} {357}},\ \bibinfo {pages} {995} (\bibinfo {year} {2017})}\BibitemShut {NoStop}%
\end{thebibliography}%

\end{document}